\documentclass[aps,pre,twocolumn,showpacs,superscriptaddress,letterpaper]{revtex4-1}

\usepackage{lmodern}
\usepackage[dvips]{graphicx}
\usepackage[normalem]{ulem}
\usepackage{upgreek,xspace}
\usepackage{psfrag,pstricks}
\usepackage{amsmath,amssymb,amsthm,graphics,mathtools}
\usepackage{siunitx}
\usepackage{float}
\usepackage[english]{babel}

\newcommand{\eq}[1]{Eq.~\eqref{#1}}
\newcommand{\Eq}[1]{Equation~\eqref{#1}}
\newcommand{\eqs}[2]{Eqs.~\eqref{#1} and~\eqref{#2}}
\newcommand{\ie}{i.e.\@\xspace}

\newcommand{\fig}[1]{Fig.~\ref{#1}}
\newcommand{\Fig}[1]{Figure~\ref{#1}}

\renewcommand{\bm}[1]{\boldsymbol{\mathbf{#1}}}

\newcommand{\grad}{\bm{\nabla}}
\newcommand{\lap}{\Delta}
\newcommand{\dd}{\mathrm{d}}

\newcommand{\dds}{\dd \Omega}

\newcommand{\im}{\operatorname{Im}}
\newcommand{\re}{\operatorname{Re}}
\newcommand{\valp}{\operatorname{VP}}
\newcommand{\dir}{\delta}

\newcommand{\ci}{j}
\newcommand{\eto}[1]{#1^*}
\newcommand{\unitv}[1]{\hat{#1} }
\newcommand{\var}{\bm{r}}

\newcommand{\varho}{\bm{\uprho}}

\newcommand{\vara}{\bm{r}}
\newcommand{\varb}{\bm{r}'}

\newcommand{\roa}{\bm{\uprho}_1}
\newcommand{\rob}{\bm{\uprho}_2}
\newcommand{\vak}{\bm{k}}

\newcommand{\vaka}{\bm{k}_1}
\newcommand{\vakb}{\bm{k}_2}

\newcommand{\vaq}{\bm{q}}

\newcommand{\vaqa}{\bm{q}_1}
\newcommand{\vaqb}{\bm{q}_2}

\newcommand{\tem}{\tau}
\newcommand{\omg}{\omega}
\newcommand{\omgo}{\Omega}
\newcommand{\opq }{\varGamma}
\newcommand{\vax}{\bm{r}}
\newcommand{\xa}{\vax_1}
\newcommand{\xb}{\vax_2}

\newcommand{\vaxs}{\bm{r}_s}
\newcommand{\vay}{\bm{\uprho}}
\newcommand{\ya}{\vay_1}
\newcommand{\yb}{\vay_2}

\newcommand{\moy}[1]{\left\langle#1\right\rangle}

\newcommand{\fours}[1]{\widetilde{#1}}
\newcommand{\fourt}[1]{\widetilde{#1}}

\newcommand{\comp}{\chi}
\newcommand{\ourco}{c_0}
\newcommand{\ourc }{c}
\newcommand{\ourpsi}{P}
\newcommand{\kr}{ k_r}
\newcommand{\ourk}{k_0}
\newcommand{\pfd}{G}

\newcommand{\opMha}{V}
\newcommand{\opMh}{\gamma}
\newcommand{\dens}{\rho}
\newcommand{\nota}{\alpha}
\newcommand{\notb}{\beta}

\newcommand{\selaa}{\Sigma_{\nota\nota}}
\newcommand{\selab}{\Sigma_{\nota\notb}}
\newcommand{\selba}{\Sigma_{\notb\nota}}
\newcommand{\selbb}{\Sigma_{\notb\notb}}
\newcommand{\sel}{\Sigma}

\newcommand{\opk}{K}
\newcommand{\opkaa}{\opk_{\alpha\alpha}}
\newcommand{\opkab}{\opk_{\alpha\beta}}
\newcommand{\opkba}{\opk_{\beta\alpha}}
\newcommand{\opkbb}{\opk_{\beta\beta}}
\newcommand{\hypa}{(\textbf{H}$_1$)\@\xspace}
\newcommand{\hypb}{(\textbf{H}$_2$)\@\xspace}
\newcommand{\hypc}{(\textbf{H}$_3$)\@\xspace}

\newcommand{\leto}{\ell^{*}}
\newcommand{\letoa}{\ell^{*(\alpha\alpha)}}
\newcommand{\ls}{\ell_s}
\newcommand{\anisfac}{g}
\newcommand{\spint}{\mathcal{I}}
\newcommand{\angkq}{\Theta}
\newcommand{\terms}{S}
\newcommand{\funph}{f}

\newcommand{\cor}{c}
\newcommand{\corac}{C_{\nota\nota}}
\newcommand{\corbc}{C_{\nota\notb}}
\newcommand{\corcc}{C_{\notb\notb}}
\newcommand{\cordc}{C_{\notb\nota}}
\newcommand{\cora}{c_{\nota\nota}}
\newcommand{\corb}{c_{\nota\notb}}
\newcommand{\corc}{c_{\notb\notb}}

\newcommand{\ctel}{\ell_c}

\newcommand{\cova }{\sigma_{\alpha}}
\newcommand{\covb }{\sigma_{\beta}}
\newcommand{\cov }{\sigma}

\newcounter{tempa}
\newcounter{tempb}
\newcounter{tempc}
\newcounter{tempd}
\newenvironment{diaga}[1]{\psset{unit=1.5mm,fillstyle=solid,fillcolor=white}
   \begin{pspicture}[shift=-2.0](0,-3)(#1,3)}{\end{pspicture}
}
\newenvironment{diagb}[1]{\psset{unit=1.5mm,fillstyle=solid,fillcolor=white}
   \begin{pspicture}[shift=-2.0](0,-3)(#1,6)}{\end{pspicture}
}

\newcommand{\ligne}[2]{\psline(#1,0)(#2,0)}

\newcommand{\particule}[1]{\pscircle(#1,0){1}}

\newcommand{\correldeux}[2]{
   \setcounter{tempa}{#2}
   \addtocounter{tempa}{-#1}
   \divide \value{tempa} by 2
   \setcounter{tempb}{#1}
   \addtocounter{tempb}{#2}
   \divide \value{tempb} by 2
   \psarc[linestyle=dashed](\value{tempb},0){\value{tempa}}{0}{180}
}

\newenvironment{ddiag}[1]{\psset{unit=1.5mm,fillstyle=solid,fillcolor=white}
   \begin{pspicture}[shift=-5](0,-6)(#1,6)}{\end{pspicture}
}

\newcommand{\pparticule}[2]{\pscircle(#1,#2){1}}

\newcommand{\iidentique}[4]{
   \psline(#1,#2)(#3,#4)
}
\newcommand{\ccorreldeuxa}[2]{
   \setcounter{tempa}{#2}
   \addtocounter{tempa}{-#1}
   \divide \value{tempa} by 2
   \setcounter{tempb}{#1}
   \addtocounter{tempb}{#2}
   \divide \value{tempb} by 2
   \psarc[linestyle=dashed](\value{tempb},3){\value{tempa}}{0}{180}
}

\newcommand{\ccorreldeuxb}[2]{
   \setcounter{tempa}{#2}
   \addtocounter{tempa}{-#1}
   \divide \value{tempa} by 2
   \setcounter{tempb}{#1}
   \addtocounter{tempb}{#2}
   \divide \value{tempb} by 2
   \psarc[linestyle=dashed](\value{tempb},-3){\value{tempa}}{180}{360}
}
\newcommand{\ccorreldeuxc}[4]{
   \psline[linestyle=dashed](#1,#2)(#3,#4)
}

\begin{document}


\title{Radiative transfer of acoustic waves in continuous complex media: Beyond the Helmholtz equation}
\author{Ibrahim Baydoun}
\author{Diego Baresch}
\author{Romain Pierrat}
\author{Arnaud Derode}
\affiliation{ESPCI Paris, PSL Research University, CNRS, Univ Paris Diderot,
Sorbonne Paris Cit\'{e}, Institut Langevin, 1 rue Jussieu, F-75005 Paris, France}
\email{arnaud.derode@espci.fr}

\date{\today}

\begin{abstract}
   Heterogeneity can be accounted for by a random potential in the wave equation. For acoustic waves in a fluid with
   fluctuations of both density and compressibility (as well as for electromagnetic waves in a medium with fluctuation
   of both permittivity and permeability) the random potential entails  a scalar and an operator contribution. For
   simplicity, the latter is usually overlooked in multiple scattering theory: whatever the type of waves, this
   simplification amounts to considering the Helmholtz equation with a sound speed $c$ depending on position $\vax$. In
   this work, a radiative transfer equation is derived from the wave equation, in order to study energy transport
   through a multiple scattering medium. In particular, the influence of the operator term on various transport
   parameters is studied, based on the diagrammatic approach of multiple scattering. Analytical results are obtained for
   fundamental quantities of transport theory such as the transport mean-free path $\leto$, scattering phase function
   $\funph$ and anisotropy factor $g$. Discarding the operator term in the wave equation is shown to have a significant
   impact on $\funph$ and $g$, yet limited to the low-frequency regime \ie,  when the correlation length of the disorder
   $\ell_c$ is smaller than or comparable to the wavelength $\lambda$. More surprisingly, discarding the operator part
   has a significant impact on the transport mean-free path $\ell^*$ whatever the frequency regime. When the scalar and
   operator terms have identical amplitudes, the discrepancy on  the transport mean-free path is around $300\%$ in the
   low-frequency regime, and still above $30\%$ for $\ell_c/\lambda=10^3$ no matter how weak fluctuations of the
   disorder are. Analytical results are supported by numerical simulations of the wave equation and Monte Carlo
   simulations.  
\end{abstract}

\maketitle

\section{Introduction}

Understanding the propagation of classical waves through strongly scattering media is of great importance for many
applications such as imaging, characterization or communication with all kinds of waves~\cite{SEBBAH-2001,
SKIPETROV-2003, Sato, Mosk2012}. 

When dealing with wave propagation a first step consists in considering an inhomogeneous medium as one particular
realization of a random process. Instead of calculating the wave field exactly for one configuration, one considers
statistical averages of the wavefield and of its intensity.  They can be determined by solving two fundamental
equations: Dyson's equation for the coherent field (\ie, the ensemble-averaged wavefield) and the Bethe-Salpether
equation for the correlation of the wavefield. Both can be derived from the wave equation, within the diagrammatic
approach of multiple scattering~\cite{FOLDY-1945,FRI68, KRT-1989, BARABANENKOV-1968-1,PSheng, Akkermans2007, VAN99}.

Once Dyson's equation is solved,  the effective phase and group velocities as well as the scattering mean-free path
$\ls$ can be determined. From a physical point of view, as the wave propagates over a distance $z$, the intensity of the
coherent part decays exponentially as $\exp(-z/\ls)$ to the benefit of the incoherent contribution. In order to
calculate the total intensity (both coherent and incoherent) it is necessary to solve the Bethe-Salpether equation. It has been long established that the Bethe-Salpether
equation can be simplified into a transport equation termed the radiative transfer equation (RTE). Further
approximations lead to an even simpler equation, the diffusion equation, which has analytical solutions and is
essentially characterized by one parameter: the diffusion constant or diffusivity $D=\leto c_{\text{tr}}/3$, with
$\leto$ the transport mean-free path and $c_{\text{tr}}$ the transport speed.

In acoustics, heterogeneity originates from space-dependence of mass density and elastic constants. From an experimental
point of view, a ballistic to diffuse transition occurs as the thickness of the sample increases. Transport parameters
($\ell_s$, $D$, $\leto$, $c_{\text{tr}}$) can be estimated from experimental measurements, for instance by studying the
average transmitted flux as a function of time~\cite{Zhang, Page1, Ramamoorthy, Weaver1995, WEA90, PAGE-1997}. A key
question is: to what extent does an estimation of the transport parameters lead to a reliable information about
structural properties of the medium (correlation lengths, variance of mass density and elastic constants)? 

In the present paper, we are interested in constructing a complete radiative transfer model for acoustic waves
propagating in a continuous but heterogeneous fluid. We focus on one particular aspect: unlike Helmholtz' equation,
where heterogeneity only appears in a space-dependence of the  sound speed, the full wave equation is taken into
account. It includes a random operator term which complicates the analysis, as will be detailed later. Discarding it, as
is usually done, amounts to considering as a starting point the Helmholtz equation, with a space-dependent sound speed
$c(\vax)$ as the only source of disorder. The question we address here is the impact of the operator term in the wave
equation on the final result, \ie the parameters that appear in the RTE and finally in the diffusion equation. When
studying energy transmission in a continuous multiple scattering medium, under which conditions is it justified to
discard the operator term in the wave equation?  

Note that radiative transfer equations can be phenomenologically established without further reference to the underlying
wave equation or medium characteristics. This is why a complete derivation of the RTE starting from the correct wave
equation is necessary here, in order to relate the transport parameters to the microstuctural properties (correlation
lengths and variances) of the heterogeneities, which is the core of the paper. Here, we deal with acoustic waves but the
exact same questions arise for transport of electromagnetic energy in a medium with fluctuations of both permeability
and permittivity, and can be addressed with the same tools~\cite{Tsang:994121}.

From a theoretical point of view, in a continuous heterogeneous fluid  without dissipation, the starting point is the
following wave equation~\cite{CHERNOV-1960,ROSS-1986} for the acoustic pressure $p\left( \vax,t\right)$:
\begin{equation}\label{EDO_t}
   \lap p-\frac{1}{c_0^2}\frac{\partial^2 p}{\partial t^2}
      =\grad\notb\left(\vax\right)\cdot\grad p
         -\frac{\nota\left(\vax\right)}{c_0^2}\frac{\partial^2 p}{\partial t^2}
\end{equation}
where $c_0$  is a reference sound speed. Note that throughout the paper, the symbols used for the wave fields (pressure
$p\left( \vax,t\right)$, velocity $\bm{v}\left( \vax,t\right)$) actually refer to the complex-valued analytic signals
associated to the real quantities.  In \eq{EDO_t}, heterogeneity arises from spatial fluctuations of two dimensionless
functions of space $\nota$ and $\notb$ which are related to mass density $\dens(\vax)$ and compressibility $\comp(\vax)$
by
\begin{align}
   \nota\left(\vax\right) & = 1- \left[\frac{\ourco}{\ourc\left(\vax\right)}\right]^2,
\\
   \notb\left(\vax\right) & = \ln\left[\frac{\dens\left(\vax\right)}{\dens_0} \right]
\end{align}
where the space-dependent sound speed is $c(\vax)=1/\sqrt{\dens(\vax)\comp(\vax)}$. $\dens_0$ is an arbitrary constant
with the dimension of a mass density. In the frequency domain (angular frequency $\omg$), the Fourier transform of
\eq{EDO_t} for the acoustic pressure $P\left(\vax,\omg\right)$ becomes
\begin{equation}\label{EDO_f}
   \lap P + \ourk^2 P   =  \ourk^2 \opMh  P.
\end{equation}
$\ourk = \omega/c_0$ and $\opMh$ is a random potential that entails both a \emph{scalar} and \emph{operator} term in the
form
\begin{equation}\label{oper-poten}
 \opMh\left(\vax\right) =\nota\left(\vax\right) +\frac{1}{\ourk^2}\grad\notb\left(\vax\right)\cdot\grad.
\end{equation}
Provided the statistical properties of $\nota$ and $\notb$, particularly their correlation functions, are known, the
ensemble-averaged (or coherent) field $\moy{P}$ can be calculated, as well as the autocorrelation $\moy{P\eto{P}}$.

It is usual to discard the operator term in \eq{oper-poten}, which greatly simplifies the calculations. This relies on
the assumption that $c(\vax)$ alone fully describes the heterogeneity of the medium, and in the following it will be
referred to as the \textit{scalar approximation}.  It is true if the mass density is constant in space. It is also true
if the mass density is not constant, as long as the compressibility is: in that case, the acoustic wave equation for the
velocity potential only involves $\alpha$ and not $\gamma$. However, when fluctuations of mass density and
compressibility coexist and have comparable amplitudes, no matter how weak they are, it results in an important error in
the scattering mean-free path $\ls$ at low frequency \ie, when the correlation length is comparable to or smaller than
the wavelength~\cite{IB-DB-RP-AD}. In that case one has to use the complete expression for the potential $\gamma$
defined in \eq{oper-poten}. The self-energy $\sel$, which is the key quantity for evaluating the average field through
the Dyson equation, can be determined using the diagrammatic approach of multiple scattering. $\sel$ contains three
additional terms due to the operator part in \eq{oper-poten}, which are not taken into account under the scalar
approximation~\cite{Jones1999,IB-DB-RP-AD,ZUCK-1995}. 

Beyond the self-energy $\sel$, here we are interested in the \textit{intensity operator} $\opk$ which is the key
quantity in the Bethe-Salpeter equation, thus driving the average intensity and correlation function of the wavefield.
Particularly we aim at evaluating the impact of the scalar approximation on $\opk$ and consequently on the transport
parameters.

The rest of the paper is organised as follows. Sections~\ref{recal} and~\ref{SectionRTE} give an overview of the
quantities and parameters that are essential to account for energy transport in random media, and how they are related
to the random potential $\gamma$ in the wave equation. In Section~\ref{exp-co-dis}, analytical solutions are calculated
in the case of an exponentially-correlated disorder. They yield expressions for the transport parameters, with or
without the operator term. In order to validate the analytical results, numerical simulations of the wave equation are
performed on an ensemble of realizations. The average transmitted energy flux is calculated as a function of time, and
compared to the solution of the RTE with the transport parameters derived analytically. In that case the RTE is solved
numerically with a Monte Carlo approach. Section~\ref{conc} concludes the paper.

\section{An overview of radiative transfer}\label{recal} 

Let us first consider the case of a homogeneous and lossless medium (reference medium). In a monochromatic regime, the
free-space Green function $G_0\left(\vax,\vaxs \right)$ is the solution of the wave equation for a point source located
at $\vaxs$:
\begin{equation}\label{helm-equ}
   \lap G_0 + \ourk^2 G_0  =\dir\left( \vax-\vaxs \right).
\end{equation}
The causal solution of \eq{helm-equ} is $G_0(\vax-\vaxs)=-\exp[jk_0|\vax-\vaxs|]/[4\pi|\vax-\vaxs|]$. For a
heterogeneous fluid, the Green function $G\left(\vax,\vaxs \right)$ associated to \eq{EDO_f} satisfies:
\begin{equation}\label{helm-equ-pot-pres}
   \lap \pfd+ \ourk^2 \pfd   =  \ourk^2 \opMh  \pfd + \dir\left( \vax-\vaxs \right) .
\end{equation}
Note that the dependency of $\pfd$ and $G_0$ on $\omg$ will only be made explicit (via the function's argument) when
different frequencies are involved. In the presence of an arbitrary distribution of sources $S(\vaxs)$ in the right-hand
side of \eq{EDO_f}, the resulting field is  
\begin{align}\label{tot-gren-p}
   \ourpsi \left( \vax,\omg\right)=\int\pfd\left(\vax,\vaxs \right) \terms\left( \vaxs\right) \dd \vaxs.
\end{align} 
The perturbed Green's function $G\left( \vax,\vaxs\right)$ can be understood as the solution of \eq{helm-equ} with an
additional source term equal to $ \ourk^2 \opMh  \pfd$, involving $G$ itself. Hence, it is usual to express it in a
recursive (Lippman-Schwinger) form:
\begin{multline}\label{Lip-Schw}
   \pfd\left(\vax,\vaxs,\omg \right) =G_0\left(\vax,\vaxs,\omg\right)
\\
   +\ourk^2 \int G_0\left(\vax,\xa,\omg\right) \opMha \left(\xa,\xb\right) \pfd\left(\xb,\vaxs ,\omg \right) \dd \xa \dd\xb,
\end{multline}
where a two-variable random potential $\opMha \left(\xa,\xb\right)$ is defined in terms of $\opMh$ as follows:
\begin{multline}\label{oper-poten-2-var}
   \opMha\left(\xa,\xb\right) =\opMh\left(\xa\right)\dir\left(\xa-\xb\right)
   =\nota\left(\xa\right)\dir\left(\xa-\xb\right)
\\
   +\frac{1}{\ourk^2}\grad\notb\left(\xa\right) \cdot\grad\dir\left(\xa-\xb\right).
\end{multline}

\subsection{Average field: Dyson's equation}

Considering $\nota$ and $\notb$ as random variables with known statistical parameters, we are now interested in
determining the ensemble average of the Green function, $\langle  G \rangle$. Iteratively substituting $G$ under the
integral on the right hand side of \eq{Lip-Schw} provides an infinite sum of integrals known as Born's expansion.
After ensemble averaging this expansion, it can be shown that $\langle G \rangle$ obeys Dyson's
equation~\cite{KRT-1989}:
\begin{multline}\label{dyson}
   \moy{\pfd\left(\vax,\vaxs \right) }=G_0\left(\vax,\vaxs\right)
\\
   +\ourk^2 \int G_0\left(\vax,\xa\right) \sel \left(\xa,\xb\right) \moy{\pfd\left(\xb,\vaxs\right)} \dd \xa \dd\xb.
\end{multline}
$\sel$ is the \emph{self-energy} or \emph{mass operator} and accounts for all orders of multiple scattering events which
cannot be factorized in the ensemble average process. Assuming that the medium is statistically homogeneous, $\opMha$,
$G_0$ and consequently $\sel$ and $\moy{\pfd}$ are invariant under translation. In that case \eq{dyson} is a double
convolution product. Therefore, its spatial Fourier transform, denoted by a tilde $\widetilde{\cdot}$ symbol, is
\begin{align}\label{green-rel}
   \moy{\fours{\pfd}\left(\vak\right)}=\frac{1}{\ourk^2 -k^2-\fours{\sel}\left(\vak,\omg \right)},
\end{align}
where $\vak$ is the dual variable for $\vax-\vaxs$.

Performing an inverse Fourier transform and taking into account the source distribution yields the coherent field 
\begin{align}\label{Pmoyen}
   \moy{\ourpsi \left( \vax,\omg\right)}=\int\moy{\pfd\left(\vax-\vaxs \right)} \terms\left( \vaxs\right) \dd \vaxs.
\end{align}
The last step in determining $\moy{G}$ would be to obtain an explicit expression of $\sel$. Unfortunately, the exact
calculation is intractable in most cases of interest. But an expression as a series of Feynman's diagrams can be derived
and has been extensively discussed in Refs~\onlinecite{FRI68} and~\onlinecite{KRT-1989}. The second-order approximation
of this series, known as the Bourret approximation, will be used in section~\ref{SectionRTE} to derive an expression of
$\sel$ valid for weakly disordered systems ($k_0\ell_s\gg 1$). A complete analysis of the coherent field's propagation
is not our present purpose and the interested reader may refer to Ref.~\onlinecite{IB-DB-RP-AD} for details.
Importantly, the intensity of the coherent field, also known as the coherent or ballistic intensity, is shown to be
spatially damped with a decay length $\ls$.  When $\alpha$ and $\beta$ have similar fluctuations, the scalar
approximation has been shown to significantly overestimate $\ls$ at low frequencies, but is reasonably valid as long as
$k_0\ell_c>10$~\cite{IB-DB-RP-AD} where $\ell_c$ is the correlation length of the disorder.

\subsection{Two-point correlation of the field: Bethe-Salpeter equation}\label{bet-sal-pa} 

The intensity of the average field only describes coherent transmission through a disordered medium, it does not suffice
to account for total energy transmission, both coherent and incoherent. To do so, since all physical quantities related
to average energy involve average products of two wavefields, the essential ingredient is the two-point correlation
function $\moy{\ourpsi \left(\vara ,\omg^+ \right)\ourpsi^* \left(\varb ,\omg^- \right)}$ where $\cdot^*$ indicates a
complex conjugation and $\omg^{\pm}$ denote angular frequencies.  It is known to obey the Bethe-Salpeter
equation~\cite{FRI68,KRT-1989,Bethe-sal-nuc} which reads
\begin{multline}\label{bet-sal} 
   \moy{\ourpsi \left(\vara ,\omg^+ \right)\ourpsi^* \left(\varb ,\omg^- \right)}
      =\moy{\ourpsi\left(\vara ,\omg^+\right)}\moy{\eto{\ourpsi}\left(\varb,\omg^-\right)}
\\
   +\int \dd\xa\dd\xb\dd\roa\dd\rob\moy{\pfd\left( \vara ,\xa ,\omg^+\right)}\moy{\eto{\pfd}\left(\varb,\xb,\omg^-\right)}
\\
   \times\opk \left( \xa, \xb, \roa, \rob ,\omg^+,\omg^-\right) \moy{\ourpsi \left(\roa ,\omg^+ \right)\ourpsi^* \left(\rob ,\omg^- \right)}.
\end{multline}
$\opk$ is termed the intensity operator (or ``irreducible vertex''). Similarly to the self-energy $\sel$ for the average
field, $\opk$ can be expressed as a perturbative expansion taking into account all orders of multiple scattering events.
It is in general represented by Feynman diagrams for convenience. In section~\ref{SectionRTE}, the first order
expansion, valid for weakly disordered systems and known as the Ladder approximation, will be used. Dropping the
$\omega^\pm$ dependency for brevity, the spatial Fourier transform of $K$ is defined by
\begin{multline}
   \fourt{\opk}\left( \vaka,\vakb,\vaqa,\vaqb \right) = 
      \int \dd \xa \dd \xb \dd\roa  \dd\rob \opk\left( \xa, \xb ,\roa,  \rob \right)
\\
   \times\exp\left[ \ci\left(- \xa\cdot \vaka + \xb\cdot \vakb +\roa\cdot \vaqa-  \rob \cdot\vaqb \right)\right].
\end{multline}
Assuming that the medium is stastistically homogeneous, ${\opk}$ is invariant under spatial translation which implies in
the Fourier space
\begin{multline}\label{oper-q}
   \fourt{\opk}\left( \vaka,\vakb,\vaqa,\vaqb \right)=
\\
   \left( 2\pi\right) ^3\dir\left(\vaka-\vakb-\vaqa+\vaqb \right) \fourt{\opq}\left( \vaka,\vakb,\vaqa,\vaqb \right).
\end{multline}  

\subsection{Radiative transfer equation}\label{rad-trans} 

In many configurations of interest, the average envelope of the wavefield $ \left\langle  \left|p(\bm r,t) \right|
\right\rangle $ varies at a time scale much larger than the oscillations of the field, and its spatial variation occur
at a characteristic scale much larger than the wavelength. This is sometimes referred to as the \textit{separation of
scales} hypothesis. Under this approximation and for weakly disordered systems, it can be shown (see App.~\ref{sec:app1}
for details), that \eq{bet-sal} can be transformed into the following transport equation, known as the Radiative
Transfer Equation (RTE)~\cite{CHANDRASEKHAR-1950,Margerin, TUR94a,APRESYAN-1996}
\begin{multline}\label{RTE}
   \left[\frac{\partial_\tau}{c_{\text{tr}}}+\unitv{\vak}\cdot\grad_{\var}\right]\spint\left(\var,\unitv{\vak},\tem,\omg\right)
      =-\underbrace{\frac{1}{\ell_e}\spint\left(\var,\unitv{\vak},\tem,\omg\right)}_{\text{loss}}\\
      +\underbrace{\frac{1}{4\pi\ell_s}\int_{4\pi}\dds_{\unitv{\vaq}}\funph\left(\unitv{\vak},\unitv{\vaq},\omg \right)
         \spint\left(\var,\unitv{\vaq},\tem,\omg\right)}_{\text{gain}}
      +\underbrace{\mathcal{S}(\var,\unitv{\vak},\tem,\omg)}_{\text{source}}.
\end{multline}

The physical quantity of interest in \eq{RTE} is the specific intensity $\spint$. Mathematically, it can be rigorously
defined as a Wigner transform of the wavefield (see App.~\ref{sec:app1}). Physically, $\spint$ may be interpreted as the
local power density per unit surface at point $\var$ and time $\tem$ flowing in the direction of the unit vector
$\unitv{\vak}$ when a quasi-monochromatic wave (central frequency $\omega/2\pi$) is emitted into a random medium. The
left-hand side of \eq{RTE} involves a Lagrangian derivative $\dd \spint / \dd\tem$, in the direction $\unitv{\vak}$ at
speed $c_{\text{tr}}$. If the medium was homogeneous the loss and gain terms would vanish, meaning that the amount of energy flowing in any direction $\unitv{\vak}$ would not change
over time unless some energy is provided by the source. Inhomogeneity (hence scattering) appears in the first two terms
of the right-hand side. The extinction term $-\spint/\ell_e$ describes power losses away from direction $\unitv{\vak}$
due to scattering between $\tem$ and $\tem+ \dd \tem$. On the contrary, the following term in \eq{RTE} describes power
gained from all directions $\unitv{\vaq}$ into $\unitv{\vak}$, due to scattering. The last term is the amount of power
per unit volume injected in the medium by the source. As a whole, \eq{RTE} describes an energy balance: variation of
$\spint$ between $\tem$ and $\tem+ \dd \tem$ is due to loss, gain and source.  The RTE has five essentiel ingredients: a
particular wavenumber $k_r$, a transport speed $c_{\text{tr}}$, an extinction length $\ell_e$, a scattering length
$\ell_s$ and a phase function $\funph$.  The latter represents the probability of sound propagating in direction
$\unitv{\vaq}$ to be scattered into the solid angle $\dds_{\unitv{\vak}}$ around $\vak$. In the detailed derivation of
the RTE, these parameters are respectively given by (see Apps.~\ref{ward} and~\ref{sec:app2})
\begin{align} \label{def-el-etr}
   \kr^2\left(\omg\right) & =\re\left[ \ourk^2 -\fours{\sel}\left(\kr,\omg \right)\right],
\\\label{def_c_tr}
c_{\text{tr}}\left(\omg\right) & =   c_0^2k_r\left(\omg\right)/\omg,
\\\label{def_ext_etr}
   \frac{1}{\ell_e\left(\omg \right)} & = -\frac{1}{\kr}\im\left[ \fours{\sel}\left(\kr,\omg \right) \right],
\\\label{def_sca_etr}
   \frac{1}{\ell_s\left(\omg \right)} & = \frac{ 1 }{16\pi^2}  \int_{4\pi}
      \fourt{\opq}\left(\kr\unitv{\vak},\kr\unitv{\vak},\kr\unitv{\vaq},\kr\unitv{\vaq},\omg,\omg\right)\dds_{\unitv{\vaq}},
\\\label{def_phase_etr}
   \funph\left(\unitv{\vak} \cdot \unitv{\vaq} ,\omg \right) & =
      \frac{\ell_s(\omg)}{4\pi}\fourt{\opq}\left(\kr\unitv{\vak},\kr\unitv{\vak},\kr\unitv{\vaq},\kr\unitv{\vaq},\omg,\omg\right).
\end{align}
In these expressions, we have used the fact that the medium is also statistically isotropic. It implies that
$\fours{\sel}(\vak)$ and $\fourt{\opq}\left(\kr\unitv{\vak},\kr\unitv{\vak},\kr\unitv{\vaq},\kr\unitv{\vaq}\right)$
depend only on $|\vak|$ and $\unitv{\vak} \cdot \unitv{\vaq}$ respectively [see \eq{oper-q}]. \Eq{def-el-etr} gives an
implicit expression for $\kr$, a quantity necessary to determine all other parameters entering the RTE. The last step to
fully determine the five coefficients [Eqs.~\eqref{def_c_tr} to \eqref{def_phase_etr}] consists in relating them to the
microscopic features of the heterogeneous medium, particularly the correlation function of the random potential in the
wave equation. This is the subject of the next section.

\section{Expressions of the RTE parameters}\label{SectionRTE} 

\subsection{First-order smoothing approximation and its relation to energy conservation}\label{sec:FOSA}

At this point, it is necessary to specify the perturbative development of $\sel$ and $\opk$ as infinite series of
scattering diagrams, in order to derive explicit expressions of the extinction ($\ell_e$) and scattering ($\ell_s$)
lengths and of the phase function $f$. With the usual conventions, the development of $\sel$ can be represented as
\begin{equation}\label{Diagrammes-dy}
   \sel=
     \begin{diaga}{2}
         \particule{1}
      \end{diaga}+
      \begin{diaga}{8}
         \correldeux{1}{7}
         \ligne{1}{7}
         \particule{1}
         \particule{7}
      \end{diaga}+
      \begin{diagb}{14}
         \correldeux{1}{13}
         \ligne{1}{7}
         \ligne{7}{13}
         \particule{1}
         \particule{7}
         \particule{13}
      \end{diagb}+\ldots
\end{equation}
In this representation, circles denote scattering events (potential $V$), horizontal solid lines represent free-space
Green functions $G_0$ and dashed lines stand for spatial correlation between points. Regarding $\opk$, we have
\begin{multline}\label{Diagrammes-bs}
\opk=
   \begin{ddiag}{2}
      \ccorreldeuxc{1}{3}{1}{-3}
      \pparticule{1}{-3}
      \pparticule{1}{3}
   \end{ddiag}
   +
   \begin{ddiag}{8}
      \ccorreldeuxa{1}{7}
      \ccorreldeuxc{1}{3}{1}{-3}
      \iidentique{1}{3}{7}{3}
      \pparticule{1}{-3}
      \pparticule{1}{3}
      \pparticule{7}{3}
   \end{ddiag}
   +      
   \begin{ddiag}{8}
      \ccorreldeuxa{1}{7}
      \ccorreldeuxc{7}{3}{7}{-3}
      \iidentique{1}{3}{7}{3}
      \pparticule{7}{-3}
      \pparticule{1}{3}
      \pparticule{7}{3}
   \end{ddiag}
   +
   \begin{ddiag}{8}
      \ccorreldeuxb{1}{7}
      \ccorreldeuxc{1}{3}{1}{-3}
      \iidentique{1}{-3}{7}{-3}
      \pparticule{1}{-3}
      \pparticule{1}{3}
      \pparticule{7}{-3}
   \end{ddiag}      
   + 
   \begin{ddiag}{8}
      \ccorreldeuxb{1}{7}
      \ccorreldeuxc{7}{3}{7}{-3}
      \iidentique{1}{-3}{7}{-3}
      \pparticule{7}{-3}
      \pparticule{1}{-3}
      \pparticule{7}{3}
   \end{ddiag}
   +\ldots
\end{multline}

The upper line represents contributions to the wave field, and the bottom line to its conjugate.  Under Bourret's
approximation, only the first two diagrams in the development of $\sel$ are kept. The first one is proportional to
$\moy{\opMha\left(\xa,\xb\right)}$: it is zero as long as the reference speed $c_0$ is chosen such that $\moy{\alpha}=0$
and the medium is statistically invariant under translation (\ie $\moy{\notb}$ does not depend on the space coordinate
$\vax$). The next diagram depends on the second-order moment of $\opMha \left(\xa,\xb\right)$. The self-energy reduces
to
\begin{multline} \label{appro-1}
   \sel\left(\xa,\xb\right) \approx  \ourk^4 \int\dd\ya\dd\yb G_0\left(\ya,\yb\right)
\\
   \times\moy{\opMha\left(\xa,\ya\right)\opMha\left(\yb,\xb\right)}.  
\end{multline}
A first-order approximation is applied to the intensity operator $K$ (Ladder
approximation). Only the first term in \eq{Diagrammes-bs} is considered, which gives
\begin{equation} \label{def-K}
   \opk\left( \xa,\xb,\ya,\yb\right)  \approx  \ourk^4\moy{\opMha\left(\xa,\ya\right)\opMha\left(\xb,\yb\right)}.
\end{equation}

From a physical point of view, care should be taken when truncating the expansions of $\sel$ and $\opk$ in order to
fulfill Ward's identity \ie, ensure energy conservation. In particular, the Bourret and Ladder approximations are not
consistent with each other unless $k_r\sim k_0$ (see App.~\ref{sec:app2}) which is a reasonable approximation in the
weak disorder limit, as will be assumed in the following. In the more general case where $k_r\ne k_0$, the general
approach presented here is still valid and ensures energy conservation, provided that one goes beyond the Bourret
approximation for $\sel$ (see App.~\ref{ward} for more details).

\subsection{Explicit expressions for $\ell_e$, $\ell_s$ and $\funph$}

The potential $\gamma$ defined in \eq{oper-poten} entails both a scalar and an operator contribution. As a consequence,
the self-energy $\Sigma$ [\eq{appro-1}] and the intensity operator $K$ [\eq{def-K}] give rise to four terms, each
involving the following correlation functions and their derivatives:
\begin{align} \label{coro-def-C}
   \begin{split}
      \corac \left(\xa,\xb\right) & = \moy{ \nota\left(\xa\right) \eto{ \nota} \left(\xb\right) },
   \\
      \corbc \left(\xa,\xb\right) & = \moy{ \nota\left(\xa\right)  \eto{\notb}  \left(\xb\right) },
   \\
      \cordc \left(\xa,\xb\right) & = \moy{ \notb\left(\xa\right)  \eto{\nota}  \left(\xb\right) },
   \\
      \corcc \left(\xa,\xb\right) & =\moy{  \notb \left(\xa\right) \eto{\notb}  \left(\xb\right)  } .
   \end{split}
\end{align}
Assuming that the medium is statistically homogeneous, the four correlation functions will solely depend on $\xa-\xb$.
Replacing $V$ in \eq{appro-1} by \eq{oper-poten-2-var} yields
\begin{equation}\label{sel-sum}
   \sel \approx \selaa+\selab+\selba+\selbb,
\end{equation}
where 
\begin{align} \label{sel-dem-mu-1}
   \begin{split}
      \selaa\left(\xa-\xb\right) & =  \ourk^4 G_0 \left(\xa-\xb\right) \corac \left(\xa-\xb\right),
   \\
      \selab\left(\xa-\xb\right) & =  -\ourk^2  \grad_{\xb}
         \cdot \left[  G_0\left(\xa-\xb\right)  \grad_{\xb} \corbc \left(\xa-\xb\right) \right],
   \\
      \selba\left(\xa-\xb\right) & = \ourk^2 \grad_{\xa} G_0 \left(\xa-\xb\right) 
         \cdot  \grad_{\xa}  \cordc \left(\xa-\xb\right),
   \\
      \selbb\left(\xa-\xb\right) & =  -\grad_{\xb} \cdot \left[ \grad_{\xb}
         \otimes  \grad_{\xa} \right.
   \\
      & \quad\left.\times \{ \corcc\left(\xa-\xb\right) \}   \grad_{\xa} G_0\left(\xa-\xb\right) \right].
  \end{split}
\end{align}
For more details on the derivation of \eq{sel-dem-mu-1} see Ref.~\onlinecite{IB-DB-RP-AD}.

As to the intensity operator, replacing $V$ in
\eq{def-K} by \eq{oper-poten-2-var} we can write $K$ as a sum of four contributions:
\begin{equation}\label{k-sum}
   \opk \approx \opkaa +\opkab+\opkba+\opkbb,
\end{equation}
where
\begin{align} \label{dev-m-k-2}
   \begin{split}
      \opkaa\left(\xa,\xb,\ya,\yb\right) & = 
         \ourk^4 \corac \left(\xa-\ya\right) \dir\left(\xa-\xb\right)
   \\ & \quad\times
         \dir\left(\ya-\yb\right),
   \\
      \opkab\left(\xa,\xb,\ya,\yb\right) & = 
         \ourk^2 \left[  \grad_{\ya}\corbc \left(\xa-\ya\right) \right.
   \\ & \quad\left.
         \cdot  \grad_{\ya} \dir\left(\ya-\yb\right) \right]
            \dir\left(\xa-\xb\right),
   \\
      \opkba\left(\xa,\xb,\ya,\yb\right) & = 
         \ourk^2 \left[ \grad_{\xa} \cordc \left(\xa-\ya\right) \right.
   \\ & \quad\left.
         \cdot  \grad_{\xa} \dir\left(\xa-\xb\right)\right]
            \dir\left(\ya-\yb\right),
   \\
      \opkbb\left(\xa,\xb,\ya,\yb\right) & =
      \grad_{\xa} \dir\left(\xa-\xb\right)
   \\ & \hspace{-0.2\linewidth}
      \cdot \left\lbrace\left[\grad_{\xa}\otimes\grad_{\ya}\corcc\left(\xa-\ya\right)\right]
         \grad_{\ya} \dir\left(\ya-\yb\right) \right\rbrace.
   \end{split}
\end{align}
Note that the scalar approximation amounts to restricting the calculation of $\sel$ and $K$ to their first term in
\eqs{sel-sum}{k-sum}. $\selaa$ and $\opkaa$ are the usual contributions to the self-energy and intensity operator as
originally given by Frisch~\cite{FRI68}.  Using \eqs{oper-q}{dev-m-k-2}, we have
\begin{equation}\label{appro-q}
   \fourt{\varGamma}\approx
      \fourt{\varGamma}_{\nota\nota}+\fourt{\varGamma}_{\nota\notb}+\fourt{\varGamma}_{\notb\nota}+\fourt{\varGamma}_{\notb\notb},
\end{equation}
where
\begin{align}
   \begin{split}
      \fourt{\varGamma}_{\alpha\alpha}\left( \vaka,\vakb,\vaqa ,\vaqb,\omg \right) &
         =\ourk^4\fourt{C}_{\alpha\alpha}(\vaka-\vakb),
   \\
      \fourt{\varGamma}_{\alpha\beta}\left( \vaka,\vakb,\vaqa ,\vaqb,\omg \right) &
         = -\ourk^2\left[\left(\vaqa -\vaqb \right) \cdot \vaqb\right] \fourt{C}_{\alpha\beta}(\vaqa -\vaqb),
   \\
      \fourt{\varGamma}_{\beta\alpha}\left( \vaka,\vakb,\vaqa ,\vaqb,\omg \right) &
         = \ourk^2\left[\left(\vaka -\vakb \right)\cdot  \vakb \right] \fourt{C}_{\beta\alpha}(\vaka-\vakb),
   \\
      \fourt{\varGamma}_{\beta\beta}\left( \vaka,\vakb,\vaqa ,\vaqb,\omg \right) &
         = \left[\left(\vaka-\vakb\right) \cdot \vakb \right] \left[\left(\vaka-\vakb\right) \cdot\vaqb \right]
   \\
      & \quad\times  \fourt{C}_{\beta\beta} \left( \vaka-\vakb \right).
   \end{split}
\end{align}
Inserting \eqs{sel-sum}{appro-q} in \eqs{def_ext_etr}{def_sca_etr}, using the fact that the correlation functions [\eq{coro-def-C}]
are real and even, and approximating $\kr$ by $\ourk$ (see App.~\ref{sec:app2} for more details), we find that the
extinction and scattering coefficients are the same (hence energy conservation in a lossless medium) and are given by
\begin{multline}\label{mu_es}
   \frac{1}{\ell_s}=\frac{1}{\ell_e}=\frac{\ourk^4}{16\pi^2} \int_{4\pi} \dds_{\unitv{\vaq}}
      \left\{\fours{C}_{\alpha\alpha}\left(\ourk \unitv{\vak} - \ourk\unitv{\vaq} \right)\right.
\\
   -\left[\left( \unitv{\vak} - \unitv{\vaq}\right) \cdot  \unitv{\vaq}\right]
           \left[\fours{C}_{\alpha\beta}\left(\ourk\unitv{\vak}-\ourk\unitv{\vaq}\right)
                +\fours{C}_{\beta\alpha}\left(\ourk\unitv{\vak}-\ourk\unitv{\vaq}\right)\right]
\\\left.
   +\left[\left( \unitv{\vak} - \unitv{\vaq}\right)\cdot  \unitv{\vaq} \right] ^2
      \fours{C}_{\beta\beta} \left(\ourk \unitv{\vak} - \ourk\unitv{\vaq} \right) \right\}.
\end{multline}
As to the phase function defined in \eq{def_phase_etr}, it is found to be
\begin{multline}\label{phase}
   \funph\left(\unitv{\vak}\cdot\unitv{\vaq} ,\omg \right)=\frac{\ourk^4\ell_s}{4\pi}
      \left\{\fours{C}_{\alpha\alpha}\left(\ourk \unitv{\vak} - \ourk\unitv{\vaq} \right)\right.
\\
   -\left[\left( \unitv{\vak} - \unitv{\vaq}\right) \cdot  \unitv{\vaq}\right]
           \left[\fours{C}_{\alpha\beta}\left(\ourk\unitv{\vak}-\ourk\unitv{\vaq}\right)
                +\fours{C}_{\beta\alpha}\left(\ourk\unitv{\vak}-\ourk\unitv{\vaq}\right)\right]
\\\left.
   +\left[\left( \unitv{\vak} - \unitv{\vaq}\right)\cdot  \unitv{\vaq} \right] ^2
      \fours{C}_{\beta\beta} \left(\ourk \unitv{\vak} - \ourk\unitv{\vaq} \right) \right\}.
\end{multline}

\section{Exponentially-correlated disorder}\label{exp-co-dis}  

Based on the Bourret and Ladder approximations, explicit expressions for all parameters involved in the RTE can be
obtained upon specification of the correlation functions defined in \eq{coro-def-C}.  In this section, we will analyze
the results obtained in the standard example of an exponentially-correlated disorder. This case has the virtue of
simplicity and allows a straightforward analysis of the importance of the operator part $\beta$ in the total potential
$\opMh$ defined in \eq{oper-poten}.

\subsection{Analytical expressions}

Let us first assume that the random processes $\alpha$ and $\beta$ are \emph{jointly} stationary and invariant under
rotation, \ie all the correlation functions defined in \eq{coro-def-C} only depend on $x=|\xa-\xb|$. In that
case, the disorder is characterized by three correlation functions
\begin{align} \label{coro-def}
   \begin{split}
      \corac \left(x\right) & = \cova^2 \cora (x),
   \\
      \corbc \left(x\right) = \cordc \left(x\right) & = \cova\covb\corb (x),
   \\
      \corcc \left(x\right) & = \covb^2\corc (x),
   \end{split}
\end{align}
where $\cova^2$ and $\covb^2$ are respectively the variances of $\nota$ and $\notb$. Making the picture even simpler, we
investigate the case where $\cova =\covb =\cov$ and $\cora=\corb=\corc=\cor$. The variance of the fluctuations $\cov^2$
appears as a multiplicative term and every parameter depends on a single correlation length $\ctel$ such that
\begin{align} \label{coro-def-exp}
   \cor(x) =\exp\left(-\frac{x}{\ctel} \right),
\end{align}   
thus,
\begin{equation}\label{C}
   \fourt{C} \left(k \right)=\sigma^2\frac{8\pi\ctel^3}{\left[1+\left( k\ctel\right) ^2 \right]^{2}}.  
\end{equation}
Analytical expressions are then found for the extinction and scattering coefficients by injecting \eq{C} into
\eq{mu_es} which leads to
\begin{multline}\label{mu_e_exp}
   \frac{\ctel}{\ell_e}=\frac{\ctel}{\ell_s}=\cov^2 \left\{\frac{1+2\left(\ourk\ctel\right)^4}
      {1+4\left(\ourk\ctel\right)^2}\right.
\\\left.
   -\frac{1-2\left(\ourk\ctel\right)^2}{4\left(\ourk\ctel\right)^2}\ln\left[1+4(\ourk\ctel)^2\right]\right\}.
\end{multline}
As a consequence of statistical invariance under rotation, the phase function in \eq{phase} depends on the unitary
vectors $\unitv{\vak}$ and $\unitv{\vaq}$ only through the angle $\Theta=\left( \unitv{\vak},\unitv{\vaq}\right)$. In
the case of an exponentially correlated disorder, we obtain
\begin{equation}\label{p_exp}
   \funph\left(\cos\Theta ,\omg \right)=\frac{2\cov^2\ell_s}{ \ctel} 
      \left[\frac{\left(\ourk\ctel\right)^2\left(2-\cos\angkq\right)}
         {1+ 2\left(\ourk\ctel\right)^2\left(1-\cos\angkq\right)}\right] ^2.
\end{equation}
If \eqs{mu_es}{phase} are restricted to their scalar contributions, different expressions are obtained for $\ell_e$,
$\ell_s$ and $f$, labelled with the superscript $(\alpha\alpha)$: 
\begin{align}\label{mu_e_exp_s}
   \frac{\ctel}{\ell_e^{(\alpha\alpha)}}=\frac{\ctel}{\ell_s^{(\alpha\alpha)}} &
      =\cov^2\frac{2\left(\ourk\ctel\right)^4}{1+4\left(\ourk\ctel\right)^{2}},
\\\label{p_exp_s}
   \funph^{(\alpha\alpha)}\left(\Theta,\omg \right) &
      =\frac{2\cov^2\ell_s^{(\alpha\alpha)}}{ \ctel}
      \left[\frac{\left(\ourk\ctel\right)^2}
         {1+2\left(\ourk\ctel\right)^2\left(1-\cos\angkq\right)}\right] ^2.
\end{align} 
Hence, the additional operator term $\beta$ is expected to have an impact on both the scattering coefficient and phase
function.  Its influence on the anisotropy factor $g$ and transport mean-free path $\leto$ is also to be examined.
$\leto$ is the typical distance beyond which the non-ballistic part of the specific intensity becomes isotropic, as if
the scattered waves had lost the memory of their initial direction. $\leto$ is related to the scattering mean-free path
$\ls$ and the phase function $f$ through
\begin{equation} \label{def-kr-long}
   \leto = \frac{\ls}{1-\anisfac}.
\end{equation}
The anisotropy factor  $g$ is the average cosine of the scattering angle:
\begin{equation}\label{g}
   \anisfac=\frac{1}{2} \int_{-1}^{1}\cos \angkq \funph\left(\cos\angkq ,\omg \right) \dd \cos \angkq.
\end{equation}
We obtain
\begin{multline} \label{g_scal_expo}
   g^{\left( \alpha\alpha\right) } = 1+\dfrac{1}{2\left( k_0\ell_c\right)^{2}}
\\
   -\left[\dfrac{1}{2\left( k_0\ell_c\right)^{2}}+\dfrac{1}{8\left( k_0\ell_c\right)^{4}}\right] \ln\left[1+4(k_0\ell_c)^2\right],
\end{multline}
 under the scalar approximation and
\begin{widetext}
   \begin{equation} \label{g_scal_expo}
      g = 1+\dfrac{3\left( k_0\ell_c\right)^{2}-2\left( k_0\ell_c\right)^{4}-20\left( k_0\ell_c\right)^{6}-\ln\left[1+4\left( k_0\ell_c\right)^2\right]
         \left[3/4+ \left( k_0\ell_c\right)^{2}-7\left( k_0\ell_c\right)^{4}+4\left( k_0\ell_c\right)^{6}\right] }
      {4\left( k_0\ell_c\right)^{4}+8\left( k_0\ell_c\right)^{8}-\ln\left[1+4\left( k_0\ell_c\right)^2\right]\left[ \left(
      k_0\ell_c\right)^{2}+2\left( k_0\ell_c\right)^{4}-8\left( k_0\ell_c\right)^{6} \right] },
   \end{equation}
\end{widetext}
when the operator term is taken into account.

\begin{figure}[!htbp]
	\centering
   \includegraphics[width=0.8\linewidth]{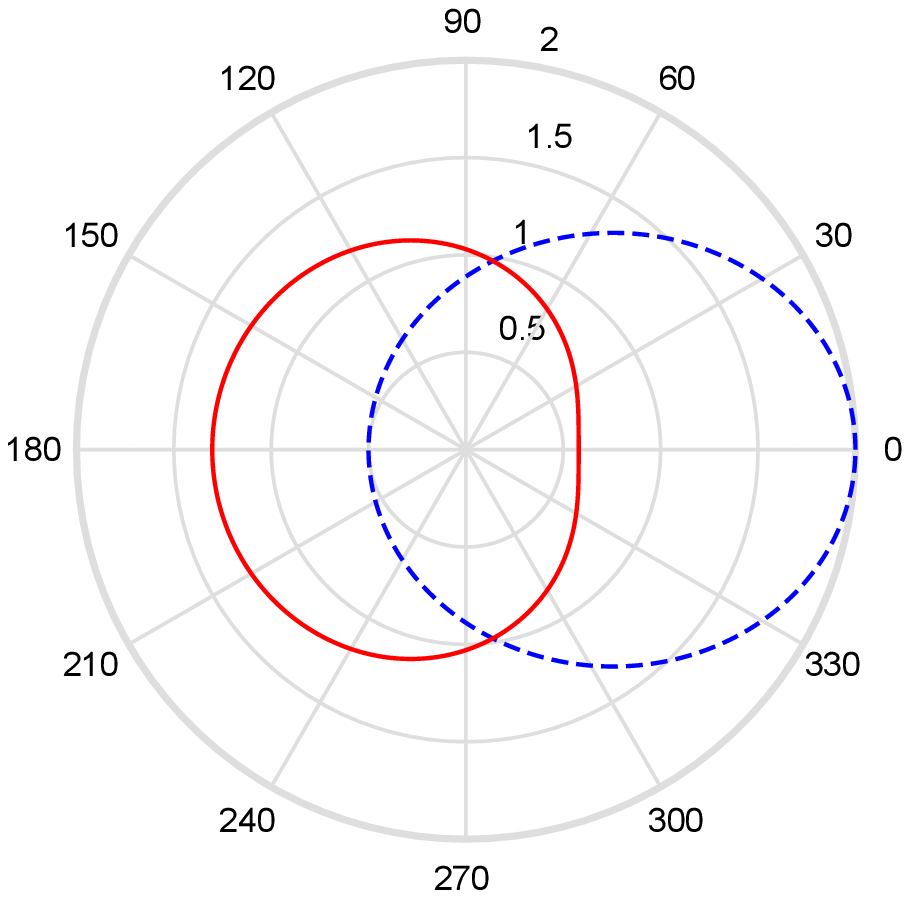}
   \\(a) $k_0\ell_c=0.5$
   \\\includegraphics[width=0.8\linewidth]{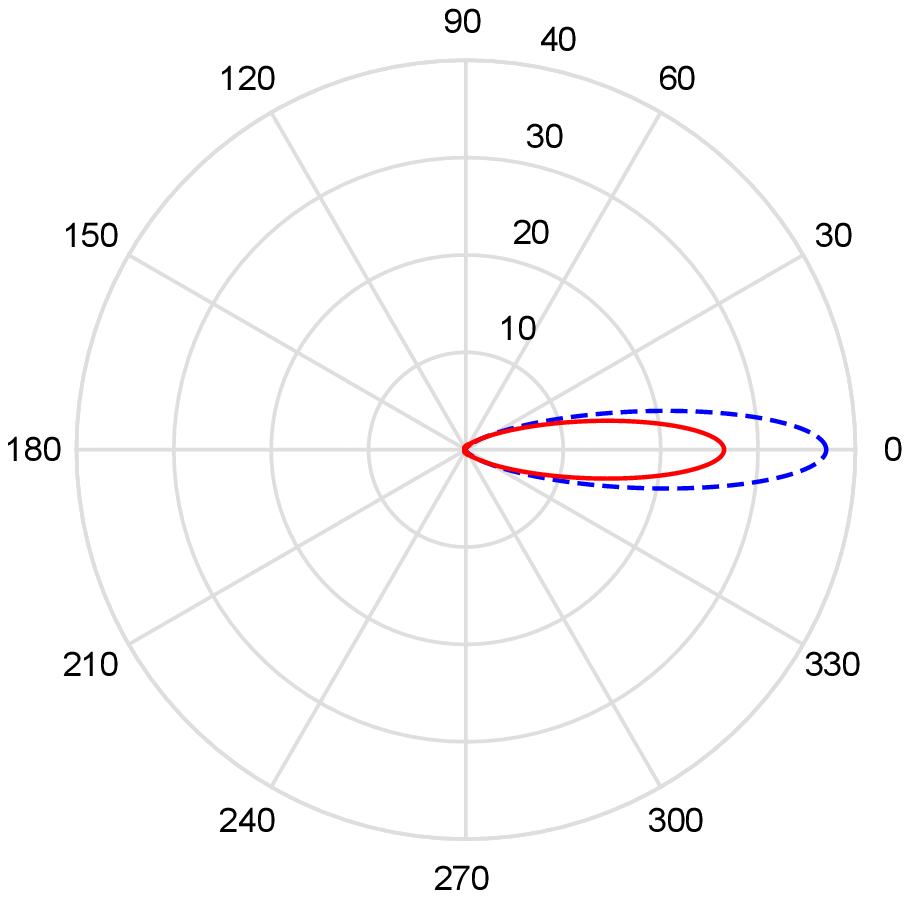}
   \\(b) $k_0\ell_c=3$
   \\\includegraphics[width=0.8\linewidth]{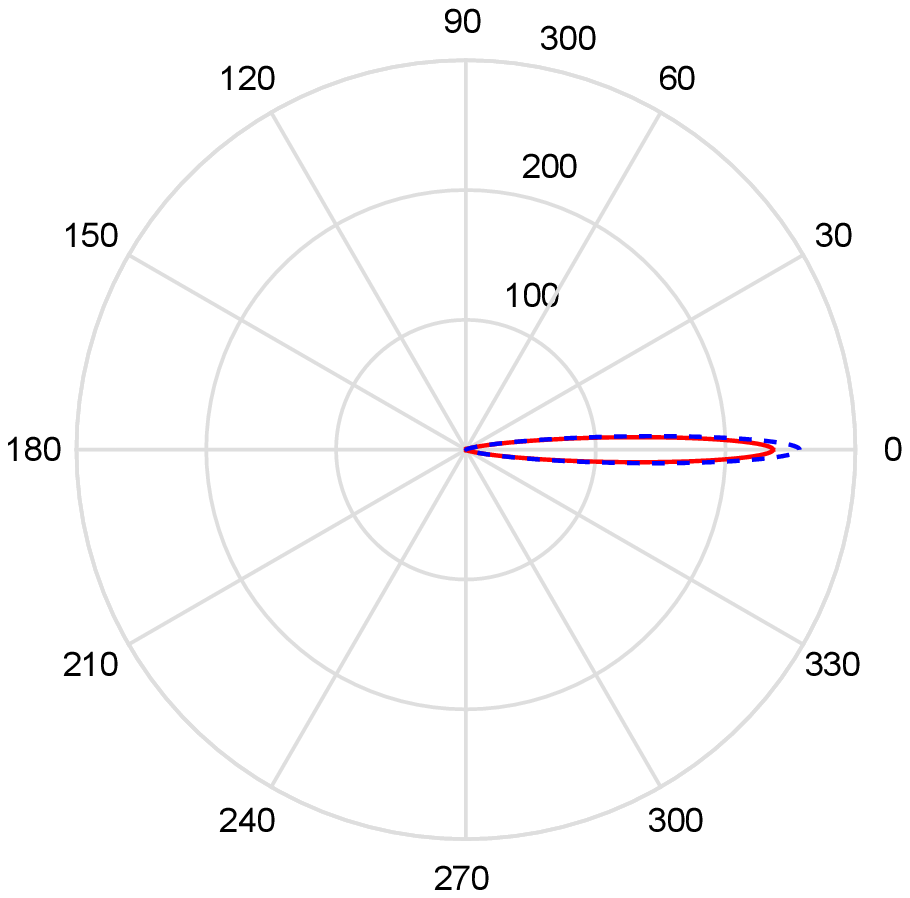}
   \\(c) $k_0\ell_c=8$
   \caption{(Color online) Polar plots of the phase functions $\funph\left(\angkq\right)$ (solid line) and
   $\funph^{(\alpha\alpha)}\left(\angkq\right)$ (dashed line) for various values of $k_0\ell_c$.}
	\label{Fig1}
\end{figure}

\begin{figure}[!htbp]
   \centering
	\psfrag{a}[c]{}
	\psfrag{x}[c]{$\ourk\ctel$}
	\psfrag{y}[c]{$\anisfac$}
   \includegraphics[width=\linewidth]{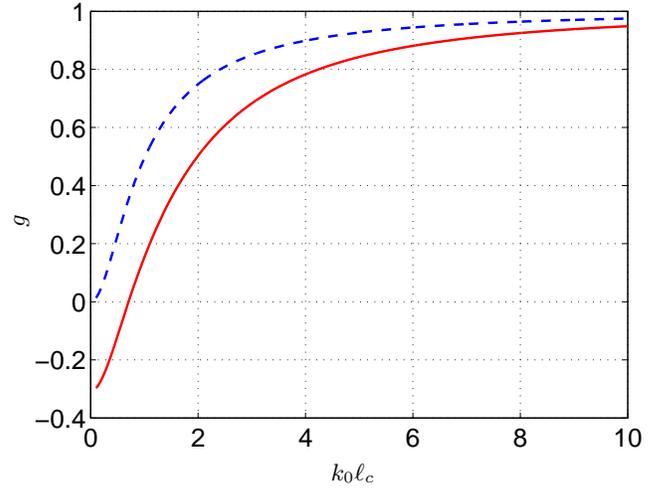}
   \caption{(Color online) Anisotropy factor $\anisfac$ as a function of $k_0\ell_c$ when the operator contribution is (solid
   line) or is not (dashed line) taken into account.}
   \label{Fig2}
\end{figure}

In order to illustrate the impact of the operator term $\beta$, the phase functions $\funph$ and
$\funph^{(\alpha\alpha)}$ are plotted in \fig{Fig1}, at three frequencies. The influence of the operator term is obvious
when the wavelength is comparable to the size of the heterogeneities.  Scattering is considerably diminished in the
forward direction. Below $k_0\ell_c = 0.7$ the anisotropy factor $g$ turns negative: the scattering pattern exhibits a
prominence to backscatter (see \fig{Fig2}). Though disorder is continuous in our case, one can draw a parallel with the
case of a homogenous medium containing discrete scattering particles. It is well known that in the low frequency
(Rayleigh) regime, the scattered pressure field is the superposition of a monopolar (omnidirectional) and a dipolar
contribution. The former is proportional to the compressibility contrast $\sigma_\chi$ between the particle and the host
fluid, while the latter is proportional to the mass density contrast $\sigma_\rho$. Depending on the amplitude and signs
of both contrasts, the resulting differential scattering-cross section exhibits a directional tendency to forward or
backward scattering. In the examples taken above, $\sigma_\alpha=\sigma_\beta$ hence compressibility and density
fluctuations are anti-correlated (in the weak fluctuations limit, we have $\sigma_\beta=\sigma_\rho$ and
$\sigma_\alpha=-\sigma_\rho-\sigma_\chi$, hence $\sigma_\beta=\sigma_\alpha$ implies $\sigma_\rho=-\sigma_\chi/2$),
which results in a prominence to backscattering in the Rayleigh regime. Note that situations for which $g<0$ also occur
for optical scatterers having both dielectric and magnetic susceptibilities~\cite{SAENZ-2012}.         

\begin{figure}[!htbp]
   \centering
   \psfrag{x}[c]{$k_0\ell_c$}
   \psfrag{y}[c]{$\sigma^2\leto/\ell_c$} 
   \includegraphics[width=\linewidth]{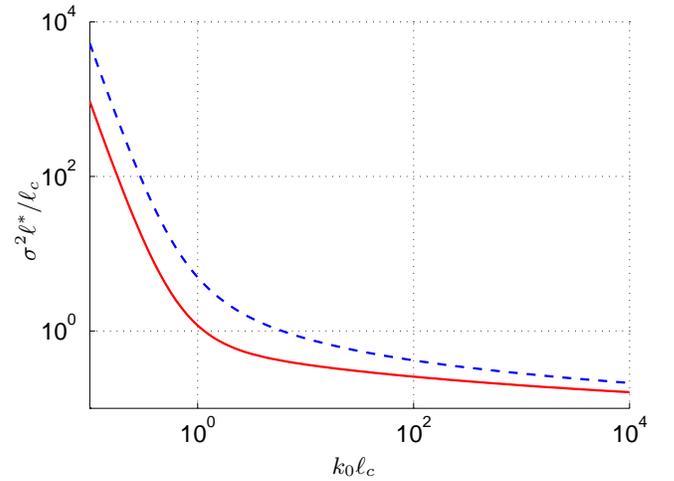}
   \caption{(Color online) Dimensionless transport mean-free path as a function of dimensionless frequency $k_0\ell_c$ when the operator contribution is (solid
   line) or is not (dashed line) taken into account.} 
   \label{Fig3}
\end{figure}

\Fig{Fig2} shows that discarding the operator term has a significant impact (larger than $\SI{10}{\%}$) on the
anisotropy factor $g$, for frequencies such that $k_0\ell_c\lesssim 5$. In the same frequency range, it has been shown
that the mean-free path $\ls$ could be nearly four times smaller than expected as $k_0\ell_c \to 0$~\cite{IB-DB-RP-AD},
whereas at higher frequencies ($k_0\ell_c>10$) the scalar approximation was reasonable.  Interestingly, this is not true
at all for the transport mean-free path. In \fig{Fig3}, $\leto$ is plotted as a function of frequency with and without
the operator contribution. The difference is far from negligible over a much broader frequency range.

In the
low-frequency regime, $1/\leto=A\sigma^2k_0^4\ell_c^3$, with $A=2$ (scalar case) or $A=34/3$ (operator case; as a consequence, $\leto$ is nearly six times smaller when the operator contribution is considered (the exact ratio is 17/3, for $k_0\ell_c\rightarrow0$). As to the anisotropy factor, $g$ vanishes in the scalar case, and is equal to $-4/13$ in the operator case. Though the results were derived in the case of an exponentially-correlated disorder, interestingly the low frequency limits for $g$, $\ell_s$ and $\leto$ do not depend on the actual shape of the correlation function (see App.~\ref{app_limit}).

At the other end of the frequency axis, $\ell_c/\leto\rightarrow\sigma^2\ln\left(k_0\ell_c \right)/2$ in both cases. But the convergence is so slow that the discrepancy persists in the high-frequency regime: it is still $\SI{30}{\%}$
for $k_0\ell_c=10^4$! The essential reason is that for transport properties, $1-g$ matters more than $g$. Even
though $g$ and $g^{\left( \alpha\alpha\right) }$ both tend to 1 (forward scattering) when $k_0\ell_c\gg1$ (see
\fig{Fig2}) the convergence is only logarithmic, hence very slow. This can be quantified by the ratio
\begin{equation}
	R=\dfrac{1-g}{1-g^{\left( \alpha\alpha\right) }}.
\end{equation}
As soon as $k_0\ell_c\gtrsim10$ the scattering mean-free paths $\ell_s$ and $\ell_s^{(\alpha\alpha)}$ are nearly the
same, so the scalar approximation is valid to evaluate the coherent field. Thus in the high-frequency regime, we have
\begin{equation}\label{R}
   \dfrac{\leto-\letoa}{\leto}=1-R=\dfrac{6}{1-2\ln\left( 2\ k_0\ell_c\right)}.
\end{equation}
If the relative error $\left| 1-R \right| $ is to be kept below $\epsilon$, it implies that $k_0\ell_c$ must be larger than $\exp\left(3/\epsilon\right)/2$. For $\epsilon=0.1$ (which would still result in a significant overestimation of the transport mean-free path), this would require $k_0\ell_c$ to be larger than $\SI{5e12}{}$, an absurdly high value from a practical point of view. Even in a situation where a high-frequency approximation ($k_0\ell_c\gg1$) seems reasonable (and in the case of $\ell_s$, the high frequency approximation does lead to $\ell_s^{(\alpha\alpha)}\approx\ell_s$ indeed as soon as $k_0\ell_c\sim10$), it is not the case for the transport mean-free path. Of course, mathematically for $k_0\ell_c\rightarrow\infty$, we have $\letoa=\leto$ as well, but how close to infinity does $k_0\ell_c$ have to be for the approximation to hold? At least $10^{12}$, which in practical terms means never.  Moreover, it should be noted the ratio of the transport mean-free paths $\leto$ and $\letoa$ only depend on $k_0\ell_c$, not on the fluctuation level $\sigma$. It implies that \textit{no matter how weak the fluctuations, the scalar approximation leads to incorrect results for the transport mean-free path, in an extremely broad frequency range}.

\subsection{Numerical validation in a cubic geometry}

Two numerical tools were used to validate the analytical calculations. On the one hand, the temporal wave equation
[\eq{EDO_t}] is solved using a finite-difference (FDTD) scheme for an ensemble of realizations with random spatial
fluctuations of density and compressibility. In this case, heterogeneity is essentially described by two parameters:
variance $\sigma^2$ and correlation length $\ell_c$. On the other hand, the Radiative Transfer Equation [\eq{RTE}] is
solved following a Monte Carlo approach. In this case, heterogeneity is accounted for by the phase function $\funph$ and
the extinction and scattering lengths $\ell_e$ and $\ell_s$. The results from both approaches are compared, in order to
validate the link between micro-structural parameters ($\sigma^2$, $\ell_c$) on the one hand and transport parameters
($\ell_e$, $\ell_s$ and $\funph$) on the other hand.

The FDTD simulations are performed using Simsonic \footnote{\texttt{www.simsonic.fr}}, a software developed in our lab
by Dr.~E.~Bossy~\cite{BOSSY-2004}. We consider a cubic domain (of length $L$) in a centered Cartesian grid $(x,y,z)$
excited by an omnidirectional point source located at $(0,0,-L/2)$. The reference (unperturbed) medium is water
($c_0=\SI{1500}{\meter\per\second}$ and $\rho_0=\SI{1000}{\kilo\gram\per\cubic\meter}$) and the emitted pulse has a
central frequency $f_c=\SI{1}{\mega\hertz}$. The mesh size was $\lambda/20$, where $\lambda$ is the corresponding
wavelength at the central frequency, to avoid significant  numerical dispersion. In order to avoid undesired
reflections, the domain is bounded by perfectly matched layers (PML). 3-D maps of the local wave speed and mass density
can be designed by the user (see Ref.~\onlinecite{IB-DB-RP-AD} for more details). The cases of a full potential $\gamma$
or its scalar limit [$\beta(r)=0$] can be studied; from a practical point of view this amounts to comparing two media
having the same sound speed at every point, but with or without mass density fluctuations.

\begin{figure}[!htbp]
   \centering
   \includegraphics[width=0.9\linewidth]{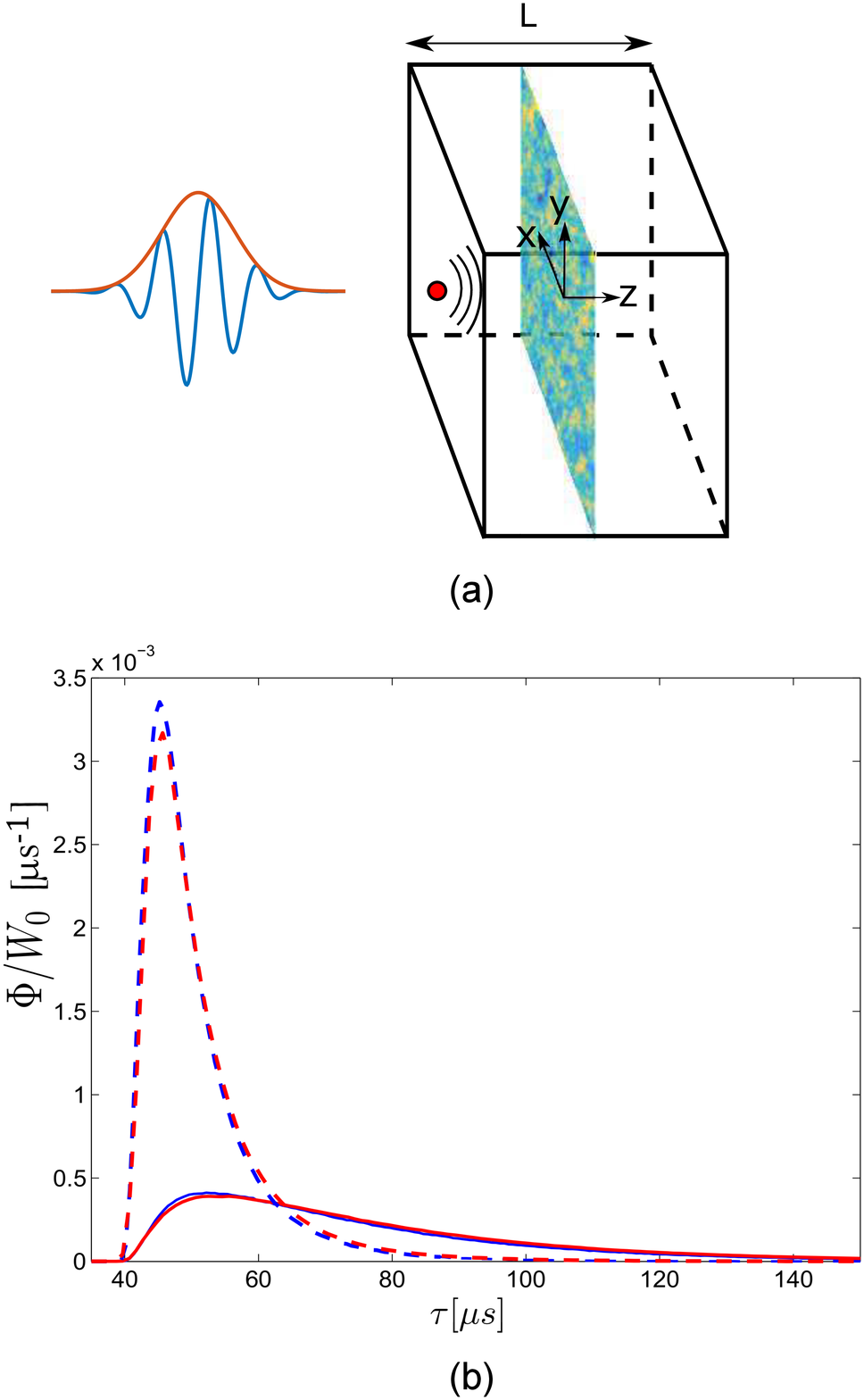}
   \caption{(Color online) Normalized acoustic flux as a function of time. (a) Sketch of the numerical experiment. The colormap
   represents the spatial fluctuations of the potentials $\alpha$ and $\beta$. A point source emits a Gaussian pulse
   with a central frequency and bandwidth of \SI{1}{\mega\hertz} on one side of a cube of length $L=118\ell_c$, with
   $\sigma=0.15$, $\ell_c\sim\SI{4.78}{\milli\meter}$ hence $\ourk \ell_c=2$. The exiting flux is measured on the
   opposite face of the cube perpendicular to $\bm{e}_z$ at $z=L/2$.  (b) Measured flux when the operator contribution
   is (solid line) or is not (dashed line) taken into account. The data are normalized by the energy $W_0$ conveyed by
   incident pulse. The blue and red curves are the flux calculated from FDTD and Monte Carlo simulations respectively.}
   \label{Flux_Sim1}
\end{figure}

A Gaussian pulse with a 1 MHz bandwidth is emitted at $t=0$ [see \fig{Flux_Sim1}(a)], its energy is denoted by $W_0$.
The real acoustic pressure $\re p(x,y,z,t)$ and particle velocity $\re \bm{v}(x,y,z,t)$ are measured at $z=L/2$.
Averaging the instantaneous Poynting vector over a period $T=1/f_c$, we have:
\begin{align}\label{Poynting} 
   \bm{J}(x,y,z,\tem) &=\frac{1}{T}\int_{\tau }^{\tau+ T} \re p(x,y,z,t)\re \bm{v}(x,y,z,t)\dd t
\\
   &=\dfrac{1}{2}\re \left( p(x,y,z,\tau)\bm{v}^*(x,y,z,\tau)\right).
\end{align}
This yields the transmitted acoustic flux $\Phi$:
\begin{equation}\label{phi_fdtd}
   \Phi(L/2,\tem)=\int_{S} \bm{J}(x,y,L/2,\tem) \cdot\dd\bm{S}.
\end{equation}
$S$ is the exit face of the cube and the infinitesimal vector $\dd\bm{S}=\mathrm{dS}\bm{e}_z$ points in the outward
direction. 

As a typical example, the normalized flux $\Phi(L/2,\tem)/W_0$ is plotted in \fig{Flux_Sim1}(b), for $\sigma=0.15$ and
$L=118\ell_c$; here $\ell_c\sim\SI{4.78}{\milli\meter}$ hence $\ourk \ell_c=2$. Interestingly, very different behaviors
are observed according to whether the operator term $\beta$ is taken into account (operator) or not (scalar). Firstly
the total transmission coefficients are \SI{1.72}{\percent} (operator), and \SI{3.84}{\percent} (scalar). Considering
\fig{Flux_Sim1}(b) as a distribution of arrival time for exiting energy packets, the average transmission times are
found to be \SI{74.9}{\micro\second} (operator) and \SI{50.6}{\micro\second} (scalar), and the standard deviations are
\SI{26.9}{\micro\second} and \SI{8.9}{\micro\second} respectively. Discarding the operator term makes the medium seem
less opaque; this is in agreement with  previous results, especially \fig{Fig3}.

In addition, a Monte Carlo simulation of the random walk of an ``acoustic particle'' (a quantum of energy $W_0$) is
performed~\cite{HAMMERSLEY-1964,FISHMAN-1996}. It is possible to show that this method can be used to solve the RTE exactly. At the
source position, an angle is picked at random with a uniform probability distribution to mimic an omnidirectional
source. The temporal profile (Gaussian envelope) is obtained by generating random departure times with a Gaussian
distribution. Once it is launched, the particle propagates in a straight line over a distance $s$. $s$ is a random
variable with probability density function $\exp(-s/\ell_s)/\ell_s$. At this stage, the phase function $\funph$ is
used to draw at random a new scattering direction. Then a new step length  $s$ is picked up and the process is
iterated. The parameters $\ls$ and $\funph$ were determined from \eqs{mu_e_exp}{p_exp} with the same variance $\sigma^2$
and correlation length $\ell_c$ as in the FDTD simulation. The random walk continues as long as the particle does not

leave the domain, then another particle is launched. The transmitted flux $\Phi(\tem)$ is incremented by
$W_0/\delta\tem$ each time a particle exits at $z=L/2$ and in the time interval $\left[ \tem,\tem+\delta\tem\right] $.

$10^8$ particles were emitted. The resulting transmitted flux is plotted in \fig{Flux_Sim1} as a function of time. The
Monte Carlo solution of the RTE are in very good agreement with the FDTD simulations of the wave equation, which
supports the analytical derivations of $\funph$ and $\ell_s$ corresponding to \eqs{mu_e_exp}{p_exp} presented earlier
and most importantly the expression of the fundamental operators $\sel$ and $\opk$ introduced in \eqs{sel-sum}{k-sum}.
The correspondance between the transmitted flux computed from either the time-averaged Poynting vector [\eq{phi_fdtd}]
or the specific intensity is established in App.~\ref{app_wigner}.

To obtain this good agreement, care must be taken to simultaneously fulfill  the criteria of validity of the different
approximations introduced in section \ref{SectionRTE}. First, it is necessary to ensure that $k_0\ell_s\gg 1$, to avoid
localization. Here, the simulations were performed for $k_0\ell_s\sim30$ and $k_0\ell_s^{(\alpha\alpha)}\sim190$.
Moreover, the Bourret and Ladder approximations require $(\sigma k_0\ell_c)^2$ to be much smaller than 1
\cite{KRT-1989}; here we took $(\sigma k_0\ell_c)^2\sim 0.09$. Finally, in order to ensure energy conservation, we must
have $k_r\simeq k_0$. In the general case, $k_r$ is defined implicitly [see \eq{def-el-etr}]. In
App.~\ref{sec:app2}, the relation between $k_r$ and $k_0$ is studied in the case of an exponential disorder. In the
scalar approximation, $k_r\simeq k_0$ always holds in the low-frequency (Rayleigh) regime. Interestingly, this is no
longer true when the operator contribution is taken into account: there is a cutoff frequency below which $k_r$
significantly deviates from $k_0$, hence the energy conservation cannot hold in the low-frequency regime, for a finite
$\sigma$. In the simulations presented here, we ensured that the condition $k_r\simeq k_0$ held, within 1 to 5 \%. If
the conditions mentioned above were not fulfilled, neither the analytical results nor the Monte Carlo solution would
match the FDTD simulations of the wave equation.

\subsection{Plane wave transmission in a slab geometry}

The operator term $\beta$ was shown to have a significant impact both on the phase function and the transport mean-free
path. In this paragraph, we study the transmission of a plane wave through an infinite slab of thickness $L$.
Considering that the analytical expressions for the transport parameters have been validated earlier, we now restrict
ourselves to the Monte Carlo simulation to calculate the transmitted flux. Indeed, for large thicknesses $L$, full
simulations of the wave equation would require much larger computational resources.

\begin{figure}[!htbp]
   \centering
   \includegraphics[width=\linewidth]{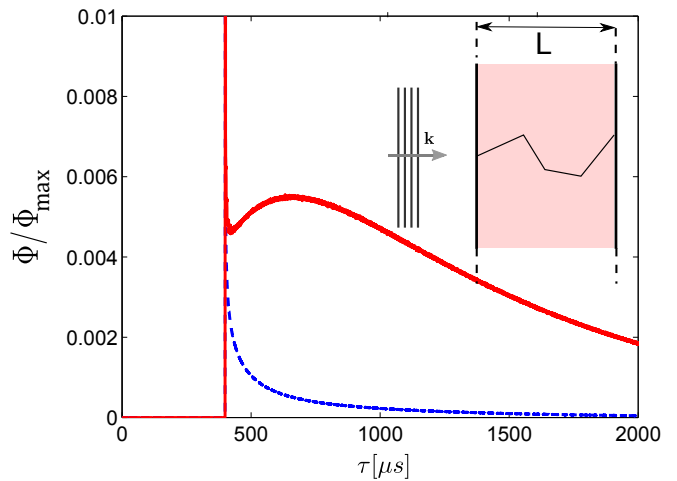}
   \caption{(Color online) Transmitted flux as a function of time in the case of an infinite slab with thickness $L=8395\ell_c$
   ($k_0\ell_c=0.3$ and $\sigma=0.1$). The flux is computed using the Monte Carlo method, when the operator part is
   taken into account (solid line) or not (dashed line). Each curve is normalized by its maximum.} 
   \label{slab1}

\end{figure}

As a first example, let us consider an infinite slab of length $L=8395\ell_c$ in the low frequency regime
($k_0\ell_c=0.3$). For $\sigma=0.1$, the sample thickness is such that $L/\ell_s^{(\alpha\alpha)}=1$  and
$L/\ell_s\simeq3.9$. In \fig{slab1}, the transmitted flux is plotted as a function of time in both cases. As can be
expected, when the operator contribution $\beta$ is dropped, wave transport is quasi-ballistic
[$L=\ell_s^{(\alpha\alpha)}$]: the sample thickness is comparable to the mean-free path, scattering events are too few to
significantly randomize the phases of the emerging waves. The ballistic arrival is found to convey 55\% of the
transmitted energy. On the contrary, if the operator contribution is considered, the transmitted intensity begins to
exhibit a diffuse coda; in that case, though the ballistic peak is still visible, it only contains 8.6\% of the
transmitted energy. 

In order to test further the diffusive nature of sound propagation, we consider a  much thicker slab. We compare the
solution of the RTE to that determined by calculating the transmitted flux using the diffusion equation. Under this
approximation, the tail of the transmitted flux decays exponentially as $\exp(-t/\tau_D)$, with 
\begin{equation}\label{tau_D}
   \tau_D=\frac{(L+2z_0)^2}{\pi^2 D}.
\end{equation}
$D=c_{\text{tr}}\leto/3$ is the diffusion constant and $z_0$ is the penetration depth beyond which sound starts to
diffuse in the sample \cite{Durian1994}. 

\begin{figure}[!htbp]
   \centering
   \psfrag{x}[c]{$\tau$ [s]}
   \psfrag{y}[c]{$\Phi/\Phi_{\text{max}}$}
   \includegraphics[width=\linewidth]{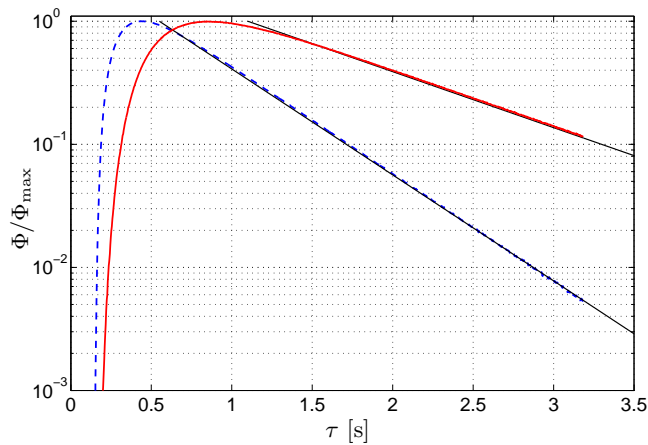}
   \caption{(Color online) Transmitted flux as a function of time in the case of an infinite slab with thickness $L=\SI{8e4}{}\ell_c$
   ($k_0\ell_c=10$ and $\sigma=0.01$). The flux is computed using the Monte Carlo method, when the operator part is
   taken into account (solid line) or not (dashed line). Each curve is normalized by its maximum. The straight lines are
   the asymptotes predicted by diffusion theory.}
   \label{slab2}
\end{figure}

\begin{figure}[!htbp]
	\centering
	\psfrag{x}[c]{$\sigma$}
	\psfrag{y}[c]{$k_0\ell_c$}
	\includegraphics[width=\linewidth]{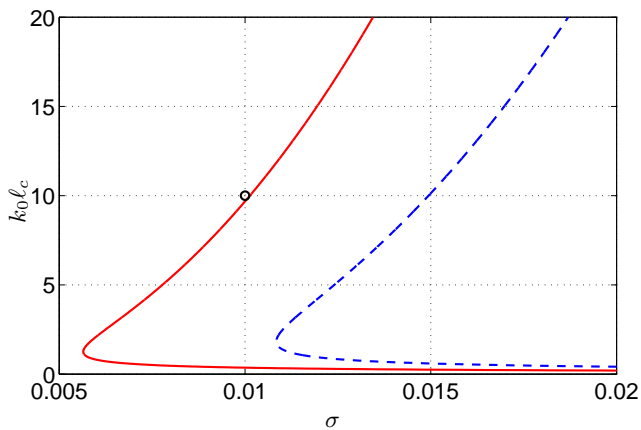}
   \caption{(Color online) Solution of the inverse problem. The pairs $(\sigma,k_0\ell_c)$ which are compatible with the value of
   $\leto$ obtained from \fig{slab2} are plotted, taking into account the operator term (continuous line) or not (dashed
   line). The exact result ($\sigma=0.01, k_0\ell_c=10)$ is represented by a circle.}
	\label{Inversion}
\end{figure}

Energy transport can safely be considered as diffusive for samples thicker than five transport mean-free paths
\cite{Zhang}. Considering medium frequency waves ($k_0\ell_c=10$) and a weak disorder ($\sigma^2=10^{-4}$) such that
$(k_0\ell_c\sigma)^2\ll1$, the transport mean-free path is expected to be $\leto=3681\ell_c$ if the operator
contribution is taken into account (operator), and $\letoa=8006 \ell_c$ under the scalar approximation (scalar). The
transmitted flux (normalized to its maximum value) is plotted in \fig{slab2} for $L=\SI{8e4}{}\ell_c$. As expected, the
diffusion approximation correctly predicts the decay time of the coda. From the slope of the tail, we obtain
$\tau_D=\SI{0.961}{\second}$ (operator) and $\tau_D=\SI{0.498}{\second}$ (scalar); the predicted values are
\SI{0.953}{\second} and \SI{0.503}{\second} respectively, assuming $z_0\simeq0.7\leto$ and $c_{\text{tr}}\simeq c_0$
\cite{Durian1994}. As a result, though the sound speed fluctuations are exactly the same in both cases the diffusion
constant $D$ varies roughly by a factor of 2.  Similarly, once $\leto$ or $D$ is measured from actual experimental data,
inverting the result to obtain a microstructural information about $\ell_c$ or $\sigma$ may result in a large mistake if
the scalar model is applied to the operator case. Here, using \eq{tau_D} we can estimate $\leto=3650~\ell_c$ from the
measured value for $\tau_D$; assuming $\ell_c$ is known, we can invert the result, and obtain
$\sigma=\SI{1.01}{\percent}$, or $\sigma=\SI{1.49}{\percent}$ under the scalar approximation (see \fig{Inversion}). The
correct value is $\sigma=\SI{1}{\percent}$ hence in this example discarding the operator term yields a \SI{50}{\percent} error on
the estimation of the fluctuations.

\section{Conclusion}\label{conc}  

In this study we have considered the transport of acoustic waves in a heterogeneous yet continuous fluid medium with
both compressibility and density fluctuations. The random potential entering the wave equation for the acoustic pressure
entails both a scalar and operator part, $\alpha$ and $\beta$. The scalar approximation consists in neglecting the
$\beta$ contribution; in that case, the space-dependent wavespeed $c(r)$ suffices to describe heterogeneity. The main
issue we addressed is the relevance of the scalar approximation when dealing with energy transport in a multiple
scattering medium. The theoretical analysis we presented is based on the diagrammatic approach of multiple scattering,
within Bourret and Ladder's approximations. The  self-energy and intensity operators $\sel$ and $\opk$ are expressed as a
function of the correlations functions of $\alpha$ and $\beta$. This relates microstructural properties (variance and
correlation lengths for $\alpha$ and $\beta$) to scattering and transport parameters. In the case of an
exponentially-correlated disorder, explicit analytical expressions are derived for the scattering and extinction
lengths, $\ell_s$ and $\ell_e$, transport speed $c_{\text{tr}}$ as well as the phase function, $f$. They are the
constitutive parameters of the radiative transfer equation (RTE) describing wave transport in scattering media.
Neglecting additional terms arising from the random operator potential $\beta$, as is usually done in the literature,
was shown to have drastic consequences on the parameters of the RTE, particularly the transport mean free path, $\leto$.  

For simplicity, we have focused on the case where the density fluctuations have a similar amplitude to that of the
compressibility (\ie same variance for $\alpha$ and $\beta$) and have an exponential correlation but the theoretical
results of \eqs{mu_es}{phase} can be applied to other cases.  In the simple case studied here, in the low frequency
regime (\ie when the wavelength is smaller or comparable to the correlation length) the operator term largely
contributes to determine the angular distribution of the reflected waves. It was also shown to have a very strong impact
on $\leto$. Its value can be down to nearly 6 times smaller than expected under the usual scalar approximation. Most
importantly the error is not restricted to a given frequency range, it persists up to the very high frequency regime
($k_0\ell_c\sim\SI{5e12}{}$): no matter how weak the fluctuations, the scalar approximation leads to incorrect results
for the transport mean-free path. The theoretical results presented here are supported by two types of numerical
simulations: FDTD simulations of the full wave equation, and Monte Carlo solution of the radiative transfer equation. 

The scattering mean-free path, the phase function, the transport mean-free path and consequently the diffusion constant
and transport speed are essential parameters to characterize wave propagation in heterogeneous media. From an
experimental point of view, they can be measured using coherent or incoherent transmission set-ups. To go beyond and
obtain a microstuctural information about the medium (fluctuations $\sigma$, correlation length $\ell_c$) one has to
invert the data with a model. Though the numerical examples were chosen to illustrate the theory in a rather academic
situation, we have shown here that if the operator term $\beta$ is ignored, the model, and consequently the estimated
values of $\sigma$ and $\ell_c$ may be completely wrong. The results presented here also open up interesting
possibilities to investigate the influence of $\beta$ on other universal wave phenomena such as coherent
backscattering.

\section*{Authorship statement}

IB and DB equally contributed to this work and share the rank of first author.

\begin{acknowledgments}
   This work was supported by the \emph{Agence Nationale de la Recherche} (ANR-11-BS09-007-01, Research Project DiAMAN),
   LABEX WIFI (Laboratory of Excellence ANR-10-LABX-24) within the French Program ``Investments for the Future'' under
   reference ANR-10-IDEX-0001-02 PSL$^{\ast}$ and by \'Electricit\'e de France R\&D.
\end{acknowledgments}

\appendix

\section{Radiative transfer equation}\label{sec:app1}

The appendix is dedicated to the derivation of the RTE from the Bethe-Salpeter equation.
Taking advantage of \eq{oper-q}, the spatial Fourier transform of \eq{bet-sal} yields
\begin{widetext}
   \begin{multline}
   \moy{
   	\fourt{\ourpsi}\left(\vak^+,\omg^+ \right)
   	\eto{\fourt{\ourpsi}}\left(\vak^-,\omg^-\right)
   }
   =\moy{\fourt{\pfd}\left(\vak^+,\omg^+\right)}
   \moy{\eto{\fourt{\pfd}}\left(\vak^-,\omg^-\right)}
   \times
   \left\{
   \fourt{\terms}\left(\vak^+ ,\omg^+ \right)
   \eto{\fourt{\terms}}\left(\vak^- ,\omg^-\right)
   \right.
   \\\left.
   +\int\frac{\dd\vaka}{\left( 2\pi\right)^3}\moy{\fourt{\ourpsi}\left(\vaka +\frac{\vaq}{2} ,\omg^+ \right)
   	\eto{\fourt{\ourpsi}}\left(\vaka -\frac{\vaq}{2} ,\omg^-\right)}
   \fourt{\opq}\left(\vak^+,\vak^-,\vaka+\frac{\vaq}{2},\vaka-\frac{\vaq}{2},\omg^+,\omg^-
   \right)\right\}.
   \end{multline}
  In the equation above we have used the notations $\bm{k}^\pm=\bm{k}\pm\bm{q}/2$,  $\bm{k}_1^\pm=\bm{k}_1\pm\bm{q}/2$
  and $\omega^\pm=\omega\pm\Omega/2$. The average Green function is expressed as 
   \begin{equation}
      \moy{\fours{\pfd}\left(\vak\right)}
         =\frac{1}{\ourk^2-k^2-\fours{\sel}\left(\vak,\omg \right)}
   \end{equation}
   and since $AB=\left(A-B\right)/\left( B^{-1}-A^{-1}\right)$ we obtain the following form of the Bethe-Salpeter equation
   \begin{multline}\label{bet-sal-3-tf}
      \left[
         \frac{\omgo\omg}{\ourco^{2}}-\vak\cdot\vaq
            -\fours{\sel}\left(\vak^+,\omg^+\right)
            +\eto{\fours{\sel}}\left(\vak^-,\omg^-\right)
      \right]
      \moy{
         \fourt{\ourpsi}\left(\vak^+,\omg^+ \right)
         \eto{\fourt{\ourpsi}}\left(\vak^-,\omg^-\right)
      }
   \\
         =\left[\moy{\fourt{\pfd}\left(\vak^+,\omg^+\right)}-
            \moy{\eto{\fourt{\pfd}}\left(\vak^-,\omg^-\right)}
         \right]
      \left\{
         \fourt{\terms}\left(\vak^+ ,\omg^+ \right)
         \eto{\fourt{\terms}}\left(\vak^- ,\omg^-\right)
      \right.
   \\\left.
      +\int\frac{\dd\vaka}{\left( 2\pi\right)^3}\moy{\fourt{\ourpsi}\left(\vaka +\frac{\vaq}{2} ,\omg^+ \right)
         \eto{\fourt{\ourpsi}}\left(\vaka -\frac{\vaq}{2} ,\omg^-\right)}
         \fourt{\opq}\left(\vak
         +\frac{\vaq}{2},\vak^-,\vaka+\frac{\vaq}{2},\vaka-\frac{\vaq}{2},\omg^+,\omg^-
         \right)\right\}.
   \end{multline}
\end{widetext}

The above equation is still exact. To derive the RTE, three assumptions are necessary:
\begin{itemize}
   \item\hypa: Separation of scales in time and space. From a physical point of view, this means that the scattered
   wavefield $p$ has a typical duration $\Delta T$ much larger than the average period $2\pi/\omega$, and a typical
   spatial extent $\Delta x$ much larger than the average wavelength $2\pi/k_0$. In other words, at any point
   (respectively, at any time) the wave field $p$ shows rapid temporal (spatial) oscillations, modulated by a slowly
   varying envelope. Reciprocally, in Fourier space, \hypa implies that $\moy{
   \fourt{\ourpsi}\left(\vak+\vaq/2),\omg+\omgo/2 \right) \eto{\fourt{\ourpsi}}\left(\vak-\vaq/2,\omg-\omgo/{2}\right)
   }$ shows the same property. The variations of $\bm{q}$ around $\bm{k}$ are limited to $\pm
   q_{\text{max}}=\pm2\pi/\Delta x\ll k $ and the variations of $\Omega$ around $\omega$ are limited to
   $\pm\Omega_{\text{max}}=\pm2\pi/\Delta T\ll \omg$.
   \item\hypb: Weak dispersion of transport parameters. The self-energy and intensity operators are supposed to vary
   slowly enough with angular frequency $\omega$ and wavenumber $k$, so that they can be considered as constant at the
   scale of $q_{\text{max}}$ and $\Omega_{\text{max}}$. 
   \item\hypc: Weak disorder assumption. It is assumed that $\im\fours{\sel}\left(\vak,\omg \right) \ll\re\left[ \ourk^2
   -\fours{\sel}\left(\vak,\omg \right)\right]$.  In a homogeneous medium, the so-called spectral function
   $\im\widetilde{G_0}$ has a singularity at $\ourk=\omega/c_0$. \hypc means that $\im\moy{\widetilde{G}}$, even if it
   does not have a true singularity, is still strongly peaked around a well-defined line in the $(\omega,\vak)$ plane,
   so that at any frequency a single effective wavenumber can be defined. In other words, the self-energy is a local
   operator.  \hypc can be interpreted as a weak fluctuation hypothesis, since it limits the allowed values  for
   $\sigma$ (see App.~\ref{sec:app2}).
\end{itemize}

If we were to perform a temporal ($\omgo\rightarrow\tem$) and spatial ($\vaq\rightarrow\var$) Fourier transform of
\eq{bet-sal-3-tf}, because of \hypa and \hypb we could do the following replacements under the Fourier integral:
\begin{align}
   \tilde{\Sigma}\left(\vak^\pm, \omg^\pm\right) & \simeq \tilde{\Sigma}\left(\vak , \omg\right),
\\\nonumber\MoveEqLeft[10]
   \fourt{\opq}\left( \vak^+,\vak^-,\vaka  +\frac{\vaq}{2}, \vaka  -\frac{\vaq}{2}, \omg^+,\omg^-\right)
\\
   & \simeq \fourt{\opq}\left(\vak,\vak,\vaka,\vaka,\omg,\omg\right),
\\ 
   \moy{\fourt{\pfd}\left(\vak^\pm,\omg^\pm\right)}
   & \simeq \moy{\fourt{\pfd}\left(\vak,\omg\right)}.
\end{align}

Moreover, in the sense of distributions, we have
\begin{equation}\label{dis-lim}
   \lim_{\varepsilon \rightarrow 0^{+}} \frac{1}{x-x_0+\ci \varepsilon} = \valp \frac{1}{x-x_0}- \ci\pi\dir\left(x-x_0 \right),
\end{equation}
where $\valp$ stands for the Cauchy principal value. Hence, using \hypc, the spectral function may be written
\begin{equation}\label{ImG}
   \im\moy{\fourt{\pfd}\left(\vak , \omg\right)}=-\pi\dir\left[k_0^2-k^2-\re\tilde{\Sigma}\left(\vak,\omg\right)\right].
\end{equation}
The Dirac delta function imposes that the modulus $k$ of the wave-vector $\vak$ must be equal to $k_r$, with
\begin{equation}
   k_r=\sqrt{\ourk^2-\re\tilde{\Sigma}\left(\kr,\omg\right)}.
\end{equation}
As a result, again, if we were to perform a temporal ($\omgo\rightarrow\tem$) and spatial ($\vaq\rightarrow\var$)
Fourier transform of \eq{bet-sal-3-tf}, the integrand could be replaced by 
\begin{widetext}
   \begin{multline}\label{bet-sal-simpl}
      \left[\frac{\omgo\omg}{\ourco^{2}}-\vak\cdot\vaq-\ci\im\fours{\sel}\left(\vak,\omg \right)\right]
     	\moy{\fourt{\ourpsi}\left(\vak+\frac{\vaq}{2},\omg+\frac{\omgo}{2} \right)
         \eto{\fourt{\ourpsi}}\left(\vak-\frac{\vaq}{2},\omg-\frac{\omgo}{2}\right)}
       =\ci\pi \dir\left[ k^2-\ourk^2-\re\tilde{\Sigma}\left(\vak,\omg\right)\right]
    \\
      \times\left\{
         \fourt{\terms}\left(\vak+\frac{\vaq}{2},\omg+\frac{\omgo}{2}\right)
            \eto{\fourt{\terms}}\left(\vak -\frac{\vaq}{2} ,\omg -\frac{\omgo}{2}\right)
      +\int\frac{\dd\vaka}{\left( 2\pi\right)^3}
         \moy{\fourt{\ourpsi}\left(\vaka +\frac{\vaq}{2} ,\omg+\frac{\omgo}{2} \right)
         \eto{\fourt{\ourpsi}}\left(\vaka -\frac{\vaq}{2} ,\omg -\frac{\omgo}{2}\right)}
      \right.
   \\
      \times\left.\vphantom{\frac{\dd\vaka}{\left( 2\pi\right)^3}}
         \fourt{\opq}\left( \vak ,\vak , \vaka ,\vaka , \omg , \omg \right)
      \right\}.
   \end{multline}
\end{widetext}

We now define a quantity $\mathcal{L}\left(\vaq,\unitv{\vak} , \omgo,\omg\right)$ such that 
\begin{multline}\label{lum}
   \frac{8\pi^3}{\kr^2 } \dir\left[k-\kr\right]\mathcal{L}\left(\vaq,\unitv{\vak} , \omgo,\omg\right)
\\
   =\moy{\fourt{\ourpsi}\left(\kr\unitv{\vak}+\frac{\vaq}{2},\omg+\frac{\omgo}{2} \right)
      \eto{\fourt{\ourpsi}}\left(\kr\unitv{\vak}-\frac{\vaq}{2},\omg-\frac{\omgo}{2}\right)}.
\end{multline} 
The temporal ($\omgo\rightarrow\tem$) and spatial ($\vaq\rightarrow\var$) Fourier transform of \eq{lum} yields
\begin{multline} \label{def_I}
   \frac{8\pi^3}{\kr^2 }\dir\left[k-\kr\right]\spint\left(\var,\unitv{\vak},\tem,\omg\right)
\\
   =\int\moy{\fourt{\ourpsi}\left(\kr\unitv{\vak}+\frac{\vaq}{2},\omg+\frac{\omgo}{2} \right)
      \eto{\fourt{\ourpsi}}\left(\kr\unitv{\vak}-\frac{\vaq}{2},\omg-\frac{\omgo}{2}\right)}
\\\times
      \exp\left[\ci\vaq\cdot\var-\ci\omgo\tem\right]\frac{\dd\vaq}{8\pi^3}\frac{\dd\omgo}{2\pi}.
\\
   =\int\moy{p\left(\var+\frac{\varho}{2},\tem+\frac{t}{2} \right)
      p^*\left(\var-\frac{\varho}{2},\tem-\frac{t}{2}\right)}
\\\times
      \exp\left[-\ci\vak\cdot\varho+\ci\omg t\right]\dd\varho\dd t.
\end{multline} 
\Eq{def_I} defines the specific intensity $\spint\left(\var,\unitv{\vak},\tem,\omg\right)$ as the spatial and temporal
Fourier transform of $\mathcal{L}$, and equivalently as the Wigner transform of the wavefield $p$. In the
phenomenological approach of RTE, $\spint$ is introduced \emph{ad hoc} as a directional decomposition (along
$\bm{\unitv{k}}$) of the power density per unit area (as a function of $\var$, $\tau$ and $\omega$), expressed in
$\si{\watt\per\square\meter\per\steradian}$, with no explicit relation to the wavefield. The Wigner transform allows a
rigorous and unambiguous mathematical definition of $\mathcal{I}$. It should be emphasized that though the Wigner
transform of the wavefield can always be defined and calculated, it can be physically interpreted as a power spectral
density only if \hypa and \hypc are valid, and $p(\var,\tem)$ denotes the complex-valued analytical signal associated to the real acoustic pressure. 

Finally, taking the spatio-temporal Fourier transform ($\omgo\rightarrow\tem, \vaq\rightarrow\var$) of
\eq{bet-sal-simpl} and inserting \eq{lum} leads to \eq{RTE}.

\section{Ward identity and energy conservation}\label{ward}

In the main text, we have shown that energy conservation is fulfilled under the Bourret and Ladder approximations for
the self energy $\sel$ and the vertex intensity $\opk$, as long as $\kr=\ourk$. In this Appendix, we show how to adapt
the Bourret approximation to ensure energy conservation even if $\kr\ne\ourk$. This derivation is adapted from
Ref.~\onlinecite{KUHN-2007-1}. We consider the most general case of a reciprocal and non-local potential
$V(\var,\varb)$. In the Ladder approximation, the vertex intensity is still given by \eq{def-K}. Using \eq{def_sca_etr},
this leads to the following expression of the scattering coefficient:
\begin{equation}
   \frac{1}{\ell_s(\omg)}=\frac{\ourk^4}{16\pi^2}
      \int\fourt{C}(\kr\unitv{\vak},\kr\unitv{\vak},\kr\unitv{\vaq},\kr\unitv{\vaq})\dd\Omega_{\unitv{\vaq}}
\end{equation}
where the correlation function $\fourt{C}$ is defined as
\begin{equation}
   \fourt{C}(\vak,\vak',\vaq,\vaq')
      =8\pi^3\dir(\vak-\vak'-\vaq+\vaq')\moy{V(\vak,\vaq)V(\vak',\vaq')}
\end{equation}
because of translational invariance (\ie statistical homogeneity of the system). Regarding the self-energy, we modify
the Bourret approximation given by \eq{appro-1} by replacing the free-space Green function by the \textit{average} one.
This leads to the following closed equation:
\begin{multline}
   \sel\left(\xa,\xb\right) \approx  \ourk^4 \int\dd\ya\dd\yb \moy{G\left(\ya,\yb\right)}
\\
   \times\moy{\opMha\left(\xa,\ya\right)\opMha\left(\yb,\xb\right)}
\end{multline}
which reads in the Fourier domain
\begin{equation}
   \fourt{\sel}\left(\vak\right)
      =\ourk^4\int\im \moy{G(\vaq)}\fourt{C}(\vak,\vak,\vaq,\vaq)\frac{\dd\vaq}{8\pi^3}.
\end{equation}
Making use of \eq{ImG}, we finally get
\begin{equation}
   \frac{1}{\ell_e(\omg)}=\frac{\ourk^4}{16\pi^2}
      \int\fourt{C}(\kr\unitv{\vak},\kr\unitv{\vak},\kr\unitv{\vaq},\kr\unitv{\vaq})\dd\Omega_{\unitv{\vaq}}
\end{equation}
which leads to $\ell_s(\omg)=\ell_e(\omg)$, hence energy conservation. In the present study, we have limited $k_0\ell_c$
and $\sigma$ to a range where $\kr\simeq\ourk$ in order to have explicit expressions for the transport parameters. Yet
it should be noted that the validity of the RTE is not restricted to this case and energy conservation can be fulfilled
even if $\kr\ne\ourk$.

\section{$k_r$ versus $k_0$}\label{sec:app2}

The particular wavenumber $k_r$ is determined by the condition
\begin{align}\label{equ-kr}
   \kr^2=\re\left[\ourk^2-\fours{\sel}\left(\kr,\omg\right)\right].
\end{align}
In the literature, it is usually assumed that $\kr \approx k_0$. This was done in Section \ref{SectionRTE} to determine
an expression for the transport coefficients of the RTE. In this appendix, we briefly show that this can be justified at
all frequencies as long as $\sigma$ is weak, under the scalar approximation. Interestingly, this is not true when the
operator contribution $\beta$ is taken into account: in that case, no matter how small $\sigma$ is, as long as it is
finite there is a cut-off frequency under which $\kr\approx k_0$ does not hold. 

In the case of an exponentially-correlated disorder and provided that the free-space Green function is used in the
Bourret approximation, it is straightforward to calculate the spatial Fourier transform of the self-energy with
\eqs{sel-sum}{sel-dem-mu-1}~\cite{IB-DB-RP-AD}. Hence two implicit and approximate expressions for \eq{equ-kr} can be
derived. In the general case, we obtain:
\begin{multline}\label{kr-op}
   \left(\kr\ctel\right)^2
      =\left(\ourk\ctel \right)^2
\\
   + \cov^2\frac{\left(\ourk\ctel\right)^4 \left[\left(\kr\ctel\right)^2+1-\left(\ourk\ctel\right)^2\right]
      -2\left(\ourk\ctel\right) ^2}
      {\left[\left(\kr\ctel\right)^2+1-\left(\ourk\ctel\right)^2\right]^2 +4\left(\ourk\ctel\right)^2}
\\
   - \cov^2\frac{1-2\left(\ourk\ctel\right)^2}{2\kr\ctel}
      \left[ \arctan\left(\ourk\ctel+\kr\ctel\right)\right.
\\
      \left.-\arctan\left(\ourk\ctel-\kr\ctel\right)\right] .
\end{multline}
And under the scalar approximation:
\begin{multline}\label{kr-sc}
   \left[\kr^{(\alpha\alpha)}\ctel \right]^2
      =\left(\ourk\ctel \right)^2
\\
   + \cov^2\frac{\left(\ourk\ctel\right)^4\left[\left(\kr\ctel\right)^2+1-\left(\ourk\ctel \right)^2\right]}
      {\left[\left(\kr\ctel\right)^2+1-\left(\ourk\ctel\right)^2\right]^2+4\left(\ourk\ctel\right)^2}.
\end{multline}

\begin{figure}[!htbp]
   \centering
	\psfrag{x}[c]{$k_0\ell_c$}
	\psfrag{y}[c]{${k_r}/{k_0}$}  
	\includegraphics[width=\linewidth]{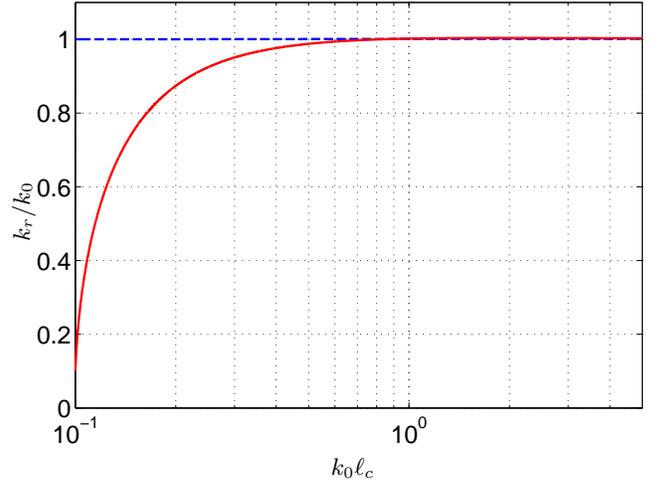}
   \caption{(Color online) Ratio between $\kr$ and $\ourk$, for $\sigma =0.1$ in the scalar (dashed line) and operator (solid line)
   cases. At this level of fluctuation, $k_0$ and $k_r$ are indistinguishable in the scalar case. On the contrary, in
   the operator case the low frequency-regime becomes clearly incompatible with the approximation $k_r\approx k_0$ as
   frequency diminishes, until \eq{kr-op} has no longer a real solution for $k_0\ell_c<0.0995$.}
	\label{kr-k0}
\end{figure}

\Eq{kr-sc} always has a real solution. In the low frequency ($k_0\ell_c\ll 1$) and low fluctuation ($\sigma \ll 1$)
approximations, it yields $k_r^{(\alpha\alpha)}=k_0(1+\sigma^2k_0^4\ell_c^4)\approx k_0$.  On the contrary for
\eq{kr-op} to have a real solution, $k_0\ell_c$ must be above a certain threshold. In the low frequency regime, with a
Taylor expansion, we find that the cut-off is approximately at $k_0\ell_c=\sigma$. For instance, with $\sigma=0.1$,
$\kr\approx k_0$ is only valid (within 5\%) if $k_0\ell_c>0.3$. The existence of a threshold and the difference between
the operator case and the scalar approximation is illustrated in \fig{kr-k0}. The low-frequency limit (Rayleigh regime)
should be handled with care in the operator case: for a finite fluctuation level $\sigma$, the $k_r\approx k_0$
approximation fails below the cut-off frequency, and the analytical expressions for transport parameters are
inapplicable. Results obtained in the Rayleigh regime are meaningful only if one makes $\sigma$ tend to zero as
$k_0\ell_c$ does.

\section{Wigner Transform and Poynting vector}\label{app_wigner}

From the specific intensity $\spint$, the average current of ``acoustic particles'' may be represented by the vector
\begin{equation}
   \bm{J}_\spint\left(\bm{r},\tau,\omega \right)=\int_{4\pi}^{} \spint\left(\bm{r},\bm{\hat{q}},\tau,\omega \right)\bm{\hat{q}}\dd\bm{\hat{q}}.
\end{equation}
Using the properties of the Dirac distribution and since $\bm{q}=q\bm{\hat{q}}$, $\bm{J}_\spint$ may be rewritten
\begin{equation}
   \bm{J}_\spint\left(\bm{r},\tau,\omega \right)
      =\dfrac{1}{k_r^3}\int_{4\pi}^{} \delta\left[ q-k_r\right] \spint\left(\bm{r},\bm{\hat{q}},\tau,\omega \right)\bm{q}\dd\bm{q}.
\end{equation}
The definition of the specific intensity [\eq{def_I}] yields
\begin{multline*}
   \bm{J}_\spint\left(\bm{r},\tau,\omega \right)=
\\
   \dfrac{1}{8\pi^3k_r}\int\moy{p\left(\var+\frac{\varho}{2},\tem+\frac{t}{2} \right)
	p^*\left(\var-\frac{\varho}{2},\tem-\frac{t}{2}\right)}
\\\times
   \exp\left[-\ci\bm{q}\cdot\varho+\ci\omg t\right] \bm{q}\dd\bm{q}\dd\varho\dd t.
\end{multline*}
Since $\bm{q}\exp\left[-\ci\bm{q}\cdot\varho\right] =\ci\grad_{\varho}\exp\left[-\ci\bm{q}\cdot\varho\right]$, an
integration by part over $\bm{\rho}$ leads to
\begin{multline*}
   \bm{J}_\spint\left(\bm{r},\tau,\omega \right)=-j\dfrac{1}{8\pi^3k_r}
\\\times
   \int\grad_{\varho}\left[ \moy{p\left(\var+\frac{\varho}{2},\tem+\frac{t}{2} \right)
	p^*\left(\var-\frac{\varho}{2},\tem-\frac{t}{2}\right)}\right]
\\\times
   \exp\left[-\ci\bm{q}\cdot\varho+\ci\omg t\right] \dd\bm{q}\dd\varho\dd t.
\end{multline*}
Besides,
\begin{equation*}
   \grad_{\varho}p\left(\var\pm\frac{\varho}{2},\tem+\frac{t}{2}
   \right)=\pm\frac{1}{2}\grad_{\var}p\left(\var\pm\frac{\varho}{2},\tem+\frac{t}{2} \right)
\end{equation*}
hence
\begin{multline*}
   \bm{J}_\spint\left(\bm{r},\tau,\omega \right)=\dfrac{1}{8\pi^3k_r}
\\\times
   \int\im\moy{\grad_{\var}\left[p\left(\var+\frac{\varho}{2},\tem+\frac{t}{2} \right)\right]
	p^*\left(\var-\frac{\varho}{2},\tem-\frac{t}{2}\right)}
\\\times
   \exp\left[-\ci\bm{q}\cdot\varho+\ci\omg t\right] \dd\bm{q}\dd\varho\dd t.
\end{multline*}
The integation over $\varho$ is straightforward, since
\begin{equation*}
   \dfrac{1}{8\pi^3}\int\exp\left[-\ci\bm{q}\cdot\varho\right] \dd\bm{q}=\delta(\varho),
\end{equation*}
hence:
\begin{multline*}
   \bm{J}_\spint\left(\bm{r},\tau,\omega \right)=\\
   \dfrac{1}{k_r}\int\im\moy{\grad_{\var}\left[p\left(\var,\tem+\frac{t}{2} \right)\right]
	p^*\left(\var,\tem-\frac{t}{2}\right)}
\\\times
   \exp\left[\ci\omg t\right] \dd\varho\dd t.
\end{multline*}
Next, we integrate over frequency and obtain
\begin{equation*}
   \int\bm{J}_\spint\left(\bm{r},\tau,\omega \right)\dd\omega=
\\
   \dfrac{2\pi}{k_r}\im\moy{\grad\left[p\left(\var,\tem \right)\right]
	p^*\left(\var,\tem\right)}.
\end{equation*}
At this stage, we can relate $\bm{J}_\spint$ to $\bm{J}$, the time-averaged acoustic Poynting vector [\eq{Poynting}]
since $\grad\left[p\left(\var,\tem \right)\right]=j\omega\rho(\var)\bm{v}\left(\var,\tem\right)$:
\begin{equation}\label{DeCadix}
   \int\bm{J}_\spint\left(\bm{r},\tau,\omega \right)\dd\omega=
\\
   \dfrac{4\pi\omega}{k_r}\moy{\rho(\var)\bm{J}\left(\var,\tem \right)}.
\end{equation}
Next, $k_r$ is approximated by $k_0$, as usual. Furthermore, in the FDTD numerical simulation, the exiting flux was
measured just behind the slab ($z=L/2^+$), in a homogenous region: in that case $\rho$ can be taken out of the bracket
in \eq{DeCadix}, to obtain
\begin{equation}
   \int\bm{J}_\spint\left(\bm{r},\tau,\omega \right)\dd\omega=
\\
   4\pi\rho_0c_0\moy{\bm{J}\left(\var,\tem\right)}.
\end{equation}
Hence, apart from a multiplicative constant with the dimensions of an acoustic impedance,  the directional average of
the specific intensity (current of ``acoustic particles'') can be identified to the frequency-averaged Poynting vector
(\si{\watt\per\square\meter}).

Note that if the exiting flux was not measured outside of the slab, the relation between $\bm{J}_\spint$ and $\bm{J}$
would only be approximate, assuming $\moy{\rho(\var)\bm{J}\left(\var,\tem \right)}\approx
\moy{\rho(\var)}\moy{\bm{J}\left(\var,\tem\right)}.$

\section{Low and high-frequency limits of $g$}\label{app_limit}

Assuming that the correlation functions $C_{\alpha\alpha}$,$C_{\alpha\beta}$ and $C_{\beta\beta}$ are identical, \eq{phase} yields
\begin{align}\label{app_phase1}
   \funph^{(\alpha\alpha)}\left(\cos\Theta,\omega\right) & =\frac{\ourk^4\ell_s^{(\alpha\alpha)}}{4\pi}
      \fours{C}\left[2k_0\left( 1-\cos\Theta\right)\right],
\\
   \funph\left(\cos\Theta,\omega\right) & =\frac{\ourk^4\ell_s}{4\pi}
      \left( 2-\cos\Theta\right)^2 \fours{C}\left[2k_0\left( 1-\cos\Theta\right)\right]. \label{app_phase2}
\end{align}
Due to the circular symmetry $\fours{C}$ can be written as a Hankel transform
\begin{equation}
   \fours{C}\left( k\right) =\frac{4\pi}{k}\int_{0}^{\infty}xC(x)\sin(kx)\dd x.
\end{equation}
Using \eq{g} and with $\mu=\cos\Theta$, we have:
\begin{multline}
	g^{(\alpha\alpha)}=\frac{k_0^3\ell_s^{(\alpha\alpha)}}{4}
      \int_{0}^{\infty}x C(x)\dd x
\\\times
      \int_{-1}^{1}\dd \mu \frac{\mu}{1-\mu} \sin\left[2k_0x(1-\mu)\right]
\end{multline}
in the scalar case, and
\begin{equation}
	g=\frac{k_0^3\ell_s}{4}\int_{0}^{\infty}x C(x)\dd x\int_{-1}^{1}\dd \mu \frac{\mu(2-\mu)^2}{1-\mu} \sin[2k_0x(1-\mu)]
\end{equation}
in the operator case.	Performing the integration over $\mu$ followed by a Taylor expansion in the low-frequency regime
$(k_0\ell_c\rightarrow 0)$ we obtain:
\begin{equation}
   g^{(\alpha\alpha)}\rightarrow\frac{4k_0^6\ell_s^{(\alpha\alpha)}}{9}\int_{0}^{\infty} x^4 C(x)\dd x
\end{equation}
and
\begin{equation}
   g\rightarrow-\frac{4k_0^4\ell_s}{3}\int_{0}^{\infty} x^2 C(x)\dd x.
\end{equation}
The low-frequency limit for $\ell_s$ is 
\begin{equation}
   \ell_s=-\frac{k_0}{\im\Sigma(k_0)}\rightarrow A \left[k_0^4\int_{0}^{\infty} x^2 C(x)\dd x\right]^{-1},
\end{equation}
with the constant $A=1$ or $A=3/13$ in the scalar and operator cases respectively~\cite{IB-DB-RP-AD}.

Hence, as $k_0\ell_c\rightarrow 0$, the anisotropy factor $g$ vanishes in the scalar case, and is equal to $-4/13$ in
the operator case. 

In the high-frequency regime, we have $\fours{C}\left( k\right)\rightarrow0$ for $k\rightarrow\infty$, while $\fours{C}(0)$ is finite and non zero. Then from \eqs{app_phase1}{app_phase2} the phase functions both tend to $0$ at all angles $\Theta$ except $\Theta=0$ (forward scattering), hence $g\rightarrow1$.

The high and low-frequency limits of $g$, $\ell_s$ and $\leto$ do not depend on the precise shape of the correlation function $C(x)$, as long as its
second and fourth moments are finite.

\bibliography{biblio}

\begin{thebibliography}{35}%
\makeatletter
\providecommand \@ifxundefined [1]{%
 \@ifx{#1\undefined}
}%
\providecommand \@ifnum [1]{%
 \ifnum #1\expandafter \@firstoftwo
 \else \expandafter \@secondoftwo
 \fi
}%
\providecommand \@ifx [1]{%
 \ifx #1\expandafter \@firstoftwo
 \else \expandafter \@secondoftwo
 \fi
}%
\providecommand \natexlab [1]{#1}%
\providecommand \enquote  [1]{``#1''}%
\providecommand \bibnamefont  [1]{#1}%
\providecommand \bibfnamefont [1]{#1}%
\providecommand \citenamefont [1]{#1}%
\providecommand \href@noop [0]{\@secondoftwo}%
\providecommand \href [0]{\begingroup \@sanitize@url \@href}%
\providecommand \@href[1]{\@@startlink{#1}\@@href}%
\providecommand \@@href[1]{\endgroup#1\@@endlink}%
\providecommand \@sanitize@url [0]{\catcode `\\12\catcode `\$12\catcode
  `\&12\catcode `\#12\catcode `\^12\catcode `\_12\catcode `\%12\relax}%
\providecommand \@@startlink[1]{}%
\providecommand \@@endlink[0]{}%
\providecommand \url  [0]{\begingroup\@sanitize@url \@url }%
\providecommand \@url [1]{\endgroup\@href {#1}{\urlprefix }}%
\providecommand \urlprefix  [0]{URL }%
\providecommand \Eprint [0]{\href }%
\providecommand \doibase [0]{http://dx.doi.org/}%
\providecommand \selectlanguage [0]{\@gobble}%
\providecommand \bibinfo  [0]{\@secondoftwo}%
\providecommand \bibfield  [0]{\@secondoftwo}%
\providecommand \translation [1]{[#1]}%
\providecommand \BibitemOpen [0]{}%
\providecommand \bibitemStop [0]{}%
\providecommand \bibitemNoStop [0]{.\EOS\space}%
\providecommand \EOS [0]{\spacefactor3000\relax}%
\providecommand \BibitemShut  [1]{\csname bibitem#1\endcsname}%
\let\auto@bib@innerbib\@empty
\bibitem [{\citenamefont {Sebbah}(2001)}]{SEBBAH-2001}%
  \BibitemOpen
  \bibinfo {editor} {\bibfnamefont {P.}~\bibnamefont {Sebbah}},\ ed.,\
  \href@noop {} {{\selectlanguage {english}\emph {\bibinfo {title} {Waves and
  Imaging through Complex Media}}}}\ (\bibinfo  {publisher} {Springer},\
  \bibinfo {address} {Dordrecht},\ \bibinfo {year} {2001})\BibitemShut
  {NoStop}%
\bibitem [{\citenamefont {van Tiggelen}\ and\ \citenamefont
  {Skipetrov}(2003)}]{SKIPETROV-2003}%
  \BibitemOpen
  \bibinfo {editor} {\bibfnamefont {B.}~\bibnamefont {van Tiggelen}}\ and\
  \bibinfo {editor} {\bibfnamefont {S.}~\bibnamefont {Skipetrov}},\ eds.,\
  \href@noop {} {{\selectlanguage {english}\emph {\bibinfo {title} {{W}ave
  Scattering in Complex Media: From Theory to Applications}}}},\ \bibinfo
  {series} {NATO Science Series II}, Vol.\ \bibinfo {volume} {107}\ (\bibinfo
  {publisher} {Springer},\ \bibinfo {address} {Dordrecht},\ \bibinfo {year}
  {2003})\BibitemShut {NoStop}%
\bibitem [{\citenamefont {Sato}\ \emph {et~al.}(2012)\citenamefont {Sato},
  \citenamefont {Fehler},\ and\ \citenamefont {Maeda}}]{Sato}%
  \BibitemOpen
  \bibfield  {author} {\bibinfo {author} {\bibfnamefont {H.}~\bibnamefont
  {Sato}}, \bibinfo {author} {\bibfnamefont {M.~C.}\ \bibnamefont {Fehler}}, \
  and\ \bibinfo {author} {\bibfnamefont {T.}~\bibnamefont {Maeda}},\
  }\href@noop {} {\emph {\bibinfo {title} {Seismic wave propagation and
  scattering in the heterogeneous earth}}},\ Vol.\ \bibinfo {volume} {484}\
  (\bibinfo  {publisher} {Springer},\ \bibinfo {year} {2012})\BibitemShut
  {NoStop}%
\bibitem [{\citenamefont {Mosk}\ \emph {et~al.}(2012)\citenamefont {Mosk},
  \citenamefont {Lagendijk}, \citenamefont {Lerosey},\ and\ \citenamefont
  {Fink}}]{Mosk2012}%
  \BibitemOpen
  \bibfield  {author} {\bibinfo {author} {\bibfnamefont {A.~P.}\ \bibnamefont
  {Mosk}}, \bibinfo {author} {\bibfnamefont {A.}~\bibnamefont {Lagendijk}},
  \bibinfo {author} {\bibfnamefont {G.}~\bibnamefont {Lerosey}}, \ and\
  \bibinfo {author} {\bibfnamefont {M.}~\bibnamefont {Fink}},\ }\href@noop {}
  {\bibfield  {journal} {\bibinfo  {journal} {Nat. Photonics}\ }\textbf
  {\bibinfo {volume} {6}},\ \bibinfo {pages} {283–292} (\bibinfo {year}
  {2012})}\BibitemShut {NoStop}%
\bibitem [{\citenamefont {Foldy}(1945)}]{FOLDY-1945}%
  \BibitemOpen
  \bibfield  {author} {\bibinfo {author} {\bibfnamefont {L.}~\bibnamefont
  {Foldy}},\ }\href@noop {} {\bibfield  {journal} {\bibinfo  {journal} {Phys.
  Rev.}\ }\textbf {\bibinfo {volume} {67}},\ \bibinfo {pages} {107} (\bibinfo
  {year} {1945})}\BibitemShut {NoStop}%
\bibitem [{\citenamefont {Frisch}(1968)}]{FRI68}%
  \BibitemOpen
  \bibfield  {author} {\bibinfo {author} {\bibfnamefont {U.}~\bibnamefont
  {Frisch}},\ }in\ \href@noop {} {\emph {\bibinfo {booktitle} {Probabilistic
  Methods in Applied Mathematics}}},\ Vol.~\bibinfo {volume} {1},\ \bibinfo
  {editor} {edited by\ \bibinfo {editor} {\bibfnamefont {A.~T.}\ \bibnamefont
  {Bharucha-Reid}}}\ (\bibinfo  {publisher} {Academic Press},\ \bibinfo
  {address} {New York, NY},\ \bibinfo {year} {1968})\ pp.\ \bibinfo {pages}
  {75--198}\BibitemShut {NoStop}%
\bibitem [{\citenamefont {Kravtsov}\ \emph {et~al.}(1989)\citenamefont
  {Kravtsov}, \citenamefont {Rytov},\ and\ \citenamefont
  {Tatarskii}}]{KRT-1989}%
  \BibitemOpen
  \bibfield  {author} {\bibinfo {author} {\bibfnamefont {Y.}~\bibnamefont
  {Kravtsov}}, \bibinfo {author} {\bibfnamefont {S.}~\bibnamefont {Rytov}}, \
  and\ \bibinfo {author} {\bibfnamefont {V.}~\bibnamefont {Tatarskii}},\
  }\href@noop {} {\emph {\bibinfo {title} {Principles Of Statistical
  Radiophysics}}}\ (\bibinfo  {publisher} {Springer-Verlag},\ \bibinfo {year}
  {1989})\BibitemShut {NoStop}%
\bibitem [{\citenamefont {Barabanenkov}(1968)}]{BARABANENKOV-1968-1}%
  \BibitemOpen
  \bibfield  {author} {\bibinfo {author} {\bibfnamefont {Y.~N.}\ \bibnamefont
  {Barabanenkov}},\ }\href@noop {} {\bibfield  {journal} {\bibinfo  {journal}
  {Sov. Phys. JETP}\ }\textbf {\bibinfo {volume} {27}},\ \bibinfo {pages} {954}
  (\bibinfo {year} {1968})}\BibitemShut {NoStop}%
\bibitem [{\citenamefont {Sheng}(2006)}]{PSheng}%
  \BibitemOpen
  \bibfield  {author} {\bibinfo {author} {\bibfnamefont {P.}~\bibnamefont
  {Sheng}},\ }\href@noop {} {\emph {\bibinfo {title} {Introduction to wave
  scattering, localization and mesoscopic phenomena}}},\ Vol.~\bibinfo {volume}
  {88}\ (\bibinfo  {publisher} {Springer Science \& Business Media},\ \bibinfo
  {year} {2006})\BibitemShut {NoStop}%
\bibitem [{\citenamefont {Akkermans}\ and\ \citenamefont
  {Montambaux}(2007)}]{Akkermans2007}%
  \BibitemOpen
  \bibfield  {author} {\bibinfo {author} {\bibfnamefont {E.}~\bibnamefont
  {Akkermans}}\ and\ \bibinfo {author} {\bibfnamefont {G.}~\bibnamefont
  {Montambaux}},\ }\href@noop {} {\emph {\bibinfo {title} {Mesoscopic physics
  of electrons and photons}}}\ (\bibinfo  {publisher} {Cambridge University
  Press},\ \bibinfo {year} {2007})\BibitemShut {NoStop}%
\bibitem [{\citenamefont {van Rossum}\ and\ \citenamefont
  {Nieuwenhuizen}(1999)}]{VAN99}%
  \BibitemOpen
  \bibfield  {author} {\bibinfo {author} {\bibfnamefont {M.~C.}\ \bibnamefont
  {van Rossum}}\ and\ \bibinfo {author} {\bibfnamefont {T.~M.}\ \bibnamefont
  {Nieuwenhuizen}},\ }\href@noop {} {\bibfield  {journal} {\bibinfo  {journal}
  {Rev. Mod. Phys.}\ }\textbf {\bibinfo {volume} {71}},\ \bibinfo {pages} {313}
  (\bibinfo {year} {1999})}\BibitemShut {NoStop}%
\bibitem [{\citenamefont {Zhang}\ \emph {et~al.}(1999)\citenamefont {Zhang},
  \citenamefont {Jones}, \citenamefont {Schriemer}, \citenamefont {Page},
  \citenamefont {Weitz},\ and\ \citenamefont {Sheng}}]{Zhang}%
  \BibitemOpen
  \bibfield  {author} {\bibinfo {author} {\bibfnamefont {Z.}~\bibnamefont
  {Zhang}}, \bibinfo {author} {\bibfnamefont {I.}~\bibnamefont {Jones}},
  \bibinfo {author} {\bibfnamefont {H.}~\bibnamefont {Schriemer}}, \bibinfo
  {author} {\bibfnamefont {J.}~\bibnamefont {Page}}, \bibinfo {author}
  {\bibfnamefont {D.}~\bibnamefont {Weitz}}, \ and\ \bibinfo {author}
  {\bibfnamefont {P.}~\bibnamefont {Sheng}},\ }\href@noop {} {\bibfield
  {journal} {\bibinfo  {journal} {Phys. Rev. E}\ }\textbf {\bibinfo {volume}
  {60}},\ \bibinfo {pages} {4843} (\bibinfo {year} {1999})}\BibitemShut
  {NoStop}%
\bibitem [{\citenamefont {Page}\ \emph {et~al.}(1995)\citenamefont {Page},
  \citenamefont {Schriemer}, \citenamefont {Bailey},\ and\ \citenamefont
  {Weitz}}]{Page1}%
  \BibitemOpen
  \bibfield  {author} {\bibinfo {author} {\bibfnamefont {J.}~\bibnamefont
  {Page}}, \bibinfo {author} {\bibfnamefont {H.}~\bibnamefont {Schriemer}},
  \bibinfo {author} {\bibfnamefont {A.}~\bibnamefont {Bailey}}, \ and\ \bibinfo
  {author} {\bibfnamefont {D.}~\bibnamefont {Weitz}},\ }\href@noop {}
  {\bibfield  {journal} {\bibinfo  {journal} {Phys. Rev. E}\ }\textbf {\bibinfo
  {volume} {52}},\ \bibinfo {pages} {3106} (\bibinfo {year}
  {1995})}\BibitemShut {NoStop}%
\bibitem [{\citenamefont {Ramamoorthy}\ \emph {et~al.}(2004)\citenamefont
  {Ramamoorthy}, \citenamefont {Kane},\ and\ \citenamefont
  {Turner}}]{Ramamoorthy}%
  \BibitemOpen
  \bibfield  {author} {\bibinfo {author} {\bibfnamefont {S.~K.}\ \bibnamefont
  {Ramamoorthy}}, \bibinfo {author} {\bibfnamefont {Y.}~\bibnamefont {Kane}}, \
  and\ \bibinfo {author} {\bibfnamefont {J.~A.}\ \bibnamefont {Turner}},\
  }\href@noop {} {\bibfield  {journal} {\bibinfo  {journal} {J. Acoust. Soc.
  Am.}\ }\textbf {\bibinfo {volume} {115}},\ \bibinfo {pages} {523} (\bibinfo
  {year} {2004})}\BibitemShut {NoStop}%
\bibitem [{\citenamefont {Weaver}\ and\ \citenamefont
  {Sachse}(1995)}]{Weaver1995}%
  \BibitemOpen
  \bibfield  {author} {\bibinfo {author} {\bibfnamefont {R.~L.}\ \bibnamefont
  {Weaver}}\ and\ \bibinfo {author} {\bibfnamefont {W.}~\bibnamefont
  {Sachse}},\ }\href@noop {} {\bibfield  {journal} {\bibinfo  {journal} {J.
  Acoust. Soc. Am.}\ }\textbf {\bibinfo {volume} {97}},\ \bibinfo {pages}
  {2094} (\bibinfo {year} {1995})}\BibitemShut {NoStop}%
\bibitem [{\citenamefont {Weaver}(1990)}]{WEA90}%
  \BibitemOpen
  \bibfield  {author} {\bibinfo {author} {\bibfnamefont {R.~L.}\ \bibnamefont
  {Weaver}},\ }\href@noop {} {\bibfield  {journal} {\bibinfo  {journal} {J.
  Mech. Phys. Solids}\ }\textbf {\bibinfo {volume} {38}},\ \bibinfo {pages}
  {55} (\bibinfo {year} {1990})}\BibitemShut {NoStop}%
\bibitem [{\citenamefont {Page}\ \emph {et~al.}(1997)\citenamefont {Page},
  \citenamefont {Schriemer}, \citenamefont {Jones}, \citenamefont {Sheng},\
  and\ \citenamefont {Weitz}}]{PAGE-1997}%
  \BibitemOpen
  \bibfield  {author} {\bibinfo {author} {\bibfnamefont {J.}~\bibnamefont
  {Page}}, \bibinfo {author} {\bibfnamefont {H.}~\bibnamefont {Schriemer}},
  \bibinfo {author} {\bibfnamefont {I.}~\bibnamefont {Jones}}, \bibinfo
  {author} {\bibfnamefont {P.}~\bibnamefont {Sheng}}, \ and\ \bibinfo {author}
  {\bibfnamefont {D.}~\bibnamefont {Weitz}},\ }\href@noop {} {\bibfield
  {journal} {\bibinfo  {journal} {Physica A}\ }\textbf {\bibinfo {volume}
  {241}},\ \bibinfo {pages} {64} (\bibinfo {year} {1997})}\BibitemShut
  {NoStop}%
\bibitem [{\citenamefont {Tsang}\ and\ \citenamefont
  {Kong}(2001)}]{Tsang:994121}%
  \BibitemOpen
  \bibfield  {author} {\bibinfo {author} {\bibfnamefont {L.}~\bibnamefont
  {Tsang}}\ and\ \bibinfo {author} {\bibfnamefont {J.~A.}\ \bibnamefont
  {Kong}},\ }\href {https://cds.cern.ch/record/994121} {\emph {\bibinfo {title}
  {Scattering of Electromagnetic Waves: advanced topics}}}\ (\bibinfo
  {publisher} {Wiley},\ \bibinfo {address} {Newark, NJ},\ \bibinfo {year}
  {2001})\BibitemShut {NoStop}%
\bibitem [{\citenamefont {Chernov}(1960)}]{CHERNOV-1960}%
  \BibitemOpen
  \bibfield  {author} {\bibinfo {author} {\bibfnamefont {L.}~\bibnamefont
  {Chernov}},\ }\href@noop {} {\emph {\bibinfo {title} {Wave Propagation in a
  Random Medium}}}\ (\bibinfo  {publisher} {McGraw Hill},\ \bibinfo {year}
  {1960})\BibitemShut {NoStop}%
\bibitem [{\citenamefont {Ross}\ and\ \citenamefont
  {Chivers}(1986)}]{ROSS-1986}%
  \BibitemOpen
  \bibfield  {author} {\bibinfo {author} {\bibfnamefont {G.}~\bibnamefont
  {Ross}}\ and\ \bibinfo {author} {\bibfnamefont {R.}~\bibnamefont {Chivers}},\
  }\href@noop {} {\bibfield  {journal} {\bibinfo  {journal} {J. Acoust. Soc.
  Am.}\ }\textbf {\bibinfo {volume} {80(5)}},\ \bibinfo {pages} {1536}
  (\bibinfo {year} {1986})}\BibitemShut {NoStop}%
\bibitem [{\citenamefont {Baydoun}\ \emph {et~al.}(2015)\citenamefont
  {Baydoun}, \citenamefont {Baresch}, \citenamefont {Pierrat},\ and\
  \citenamefont {Derode}}]{IB-DB-RP-AD}%
  \BibitemOpen
  \bibfield  {author} {\bibinfo {author} {\bibfnamefont {I.}~\bibnamefont
  {Baydoun}}, \bibinfo {author} {\bibfnamefont {D.}~\bibnamefont {Baresch}},
  \bibinfo {author} {\bibfnamefont {R.}~\bibnamefont {Pierrat}}, \ and\
  \bibinfo {author} {\bibfnamefont {A.}~\bibnamefont {Derode}},\ }\href@noop {}
  {\bibfield  {journal} {\bibinfo  {journal} {Phys. Rev. E}\ }\textbf {\bibinfo
  {volume} {92}} (\bibinfo {year} {2015})}\BibitemShut {NoStop}%
\bibitem [{\citenamefont {Jones}(1999)}]{Jones1999}%
  \BibitemOpen
  \bibfield  {author} {\bibinfo {author} {\bibfnamefont {C.}~\bibnamefont
  {Jones}},\ }\href@noop {} {\emph {\bibinfo {title} {High frequency acoustic
  volume scattering from biologically active marine sediments}}}\ (\bibinfo
  {publisher} {PhD Thesis, University of Washington},\ \bibinfo {year}
  {1999})\BibitemShut {NoStop}%
\bibitem [{\citenamefont {Zhuck}(1995)}]{ZUCK-1995}%
  \BibitemOpen
  \bibfield  {author} {\bibinfo {author} {\bibfnamefont {N.}~\bibnamefont
  {Zhuck}},\ }\href@noop {} {\bibfield  {journal} {\bibinfo  {journal} {Phys.
  Rev. B}\ }\textbf {\bibinfo {volume} {52(2)}},\ \bibinfo {pages} {919}
  (\bibinfo {year} {1995})}\BibitemShut {NoStop}%
\bibitem [{\citenamefont {Anisovich}\ \emph {et~al.}(1993)\citenamefont
  {Anisovich}, \citenamefont {Melikhov}, \citenamefont {Metsch},\ and\
  \citenamefont {Petry}}]{Bethe-sal-nuc}%
  \BibitemOpen
  \bibfield  {author} {\bibinfo {author} {\bibfnamefont {V.}~\bibnamefont
  {Anisovich}}, \bibinfo {author} {\bibfnamefont {D.}~\bibnamefont {Melikhov}},
  \bibinfo {author} {\bibfnamefont {B.}~\bibnamefont {Metsch}}, \ and\ \bibinfo
  {author} {\bibfnamefont {H.}~\bibnamefont {Petry}},\ }\href {\doibase
  10.1016/0375-9474(93)90055-3} {\bibfield  {journal} {\bibinfo  {journal}
  {Nucl. Phys. A}\ }\textbf {\bibinfo {volume} {563}},\ \bibinfo {pages} {549}
  (\bibinfo {year} {1993})}\BibitemShut {NoStop}%
\bibitem [{\citenamefont {Chandrasekhar}(1950)}]{CHANDRASEKHAR-1950}%
  \BibitemOpen
  \bibfield  {author} {\bibinfo {author} {\bibfnamefont {S.}~\bibnamefont
  {Chandrasekhar}},\ }\href@noop {} {{\selectlanguage {english}\emph {\bibinfo
  {title} {Radiative Transfer}}}}\ (\bibinfo  {publisher} {Dover},\ \bibinfo
  {address} {New-York},\ \bibinfo {year} {1950})\BibitemShut {NoStop}%
\bibitem [{\citenamefont {Margerin}(2005)}]{Margerin}%
  \BibitemOpen
  \bibfield  {author} {\bibinfo {author} {\bibfnamefont {L.}~\bibnamefont
  {Margerin}},\ }\href@noop {} {\bibfield  {journal} {\bibinfo  {journal}
  {Geophysical Monograph-American Geophysical Union}\ }\textbf {\bibinfo
  {volume} {157}},\ \bibinfo {pages} {229} (\bibinfo {year}
  {2005})}\BibitemShut {NoStop}%
\bibitem [{\citenamefont {Turner}\ and\ \citenamefont {Weaver}(1994)}]{TUR94a}%
  \BibitemOpen
  \bibfield  {author} {\bibinfo {author} {\bibfnamefont {J.~A.}\ \bibnamefont
  {Turner}}\ and\ \bibinfo {author} {\bibfnamefont {R.~L.}\ \bibnamefont
  {Weaver}},\ }\href@noop {} {\bibfield  {journal} {\bibinfo  {journal} {J.
  Acoust. Soc. Am.}\ }\textbf {\bibinfo {volume} {96}},\ \bibinfo {pages}
  {3654} (\bibinfo {year} {1994})}\BibitemShut {NoStop}%
\bibitem [{\citenamefont {Apresyan}\ and\ \citenamefont
  {Kravtsov}(1996)}]{APRESYAN-1996}%
  \BibitemOpen
  \bibfield  {author} {\bibinfo {author} {\bibfnamefont {L.~A.}\ \bibnamefont
  {Apresyan}}\ and\ \bibinfo {author} {\bibfnamefont {Y.~A.}\ \bibnamefont
  {Kravtsov}},\ }\href@noop {} {{\selectlanguage {english}\emph {\bibinfo
  {title} {Radiation Transfer: Statistical and Wave Aspects}}}}\ (\bibinfo
  {publisher} {Gordon and Breach Publishers},\ \bibinfo {address} {Amsterdam},\
  \bibinfo {year} {1996})\BibitemShut {NoStop}%
\bibitem [{\citenamefont {G{\'o}mez-Medina}\ \emph {et~al.}(2012)\citenamefont
  {G{\'o}mez-Medina}, \citenamefont {Froufe-P{\'e}rez}, \citenamefont
  {Y{\'e}pez}, \citenamefont {Scheffold}, \citenamefont {Nieto-Vesperinas},\
  and\ \citenamefont {S{\'a}enz}}]{SAENZ-2012}%
  \BibitemOpen
  \bibfield  {author} {\bibinfo {author} {\bibfnamefont {R.}~\bibnamefont
  {G{\'o}mez-Medina}}, \bibinfo {author} {\bibfnamefont {L.~S.}\ \bibnamefont
  {Froufe-P{\'e}rez}}, \bibinfo {author} {\bibfnamefont {M.}~\bibnamefont
  {Y{\'e}pez}}, \bibinfo {author} {\bibfnamefont {F.}~\bibnamefont
  {Scheffold}}, \bibinfo {author} {\bibfnamefont {M.}~\bibnamefont
  {Nieto-Vesperinas}}, \ and\ \bibinfo {author} {\bibfnamefont {J.~J.}\
  \bibnamefont {S{\'a}enz}},\ }\href {\doibase 10.1103/PhysRevA.85.035802}
  {\bibfield  {journal} {\bibinfo  {journal} {Phys. Rev. A}\ }\textbf {\bibinfo
  {volume} {85}},\ \bibinfo {pages} {035802} (\bibinfo {year}
  {2012})}\BibitemShut {NoStop}%
\bibitem [{Note1()}]{Note1}%
  \BibitemOpen
  \bibinfo {note} {\protect \texttt {www.simsonic.fr}}\BibitemShut {NoStop}%
\bibitem [{\citenamefont {Bossy}\ \emph {et~al.}(2004)\citenamefont {Bossy},
  \citenamefont {Talmant},\ and\ \citenamefont {Laugier}}]{BOSSY-2004}%
  \BibitemOpen
  \bibfield  {author} {\bibinfo {author} {\bibfnamefont {E.}~\bibnamefont
  {Bossy}}, \bibinfo {author} {\bibfnamefont {M.}~\bibnamefont {Talmant}}, \
  and\ \bibinfo {author} {\bibfnamefont {P.}~\bibnamefont {Laugier}},\
  }\href@noop {} {\bibfield  {journal} {\bibinfo  {journal} {J. Acoust. Soc.
  Am.}\ }\textbf {\bibinfo {volume} {115}},\ \bibinfo {pages} {2314} (\bibinfo
  {year} {2004})}\BibitemShut {NoStop}%
\bibitem [{\citenamefont {Hammersley}\ and\ \citenamefont
  {Handscomb}(1964)}]{HAMMERSLEY-1964}%
  \BibitemOpen
  \bibfield  {author} {\bibinfo {author} {\bibfnamefont {J.~M.}\ \bibnamefont
  {Hammersley}}\ and\ \bibinfo {author} {\bibfnamefont {D.~C.}\ \bibnamefont
  {Handscomb}},\ }\href@noop {} {{\selectlanguage {english}\emph {\bibinfo
  {title} {Monte Carlo Methods}}}}\ (\bibinfo  {publisher} {Chapman and Hall},\
  \bibinfo {address} {London},\ \bibinfo {year} {1964})\BibitemShut {NoStop}%
\bibitem [{\citenamefont {Fishman}(1996)}]{FISHMAN-1996}%
  \BibitemOpen
  \bibfield  {author} {\bibinfo {author} {\bibfnamefont {G.~S.}\ \bibnamefont
  {Fishman}},\ }\href@noop {} {{\selectlanguage {english}\emph {\bibinfo
  {title} {Monte Carlo Concepts, Algorithms and Applications}}}}\ (\bibinfo
  {publisher} {Springer Verlag},\ \bibinfo {address} {Berlin},\ \bibinfo {year}
  {1996})\BibitemShut {NoStop}%
\bibitem [{\citenamefont {Durian}(1994)}]{Durian1994}%
  \BibitemOpen
  \bibfield  {author} {\bibinfo {author} {\bibfnamefont {D.~J.}\ \bibnamefont
  {Durian}},\ }\href@noop {} {\bibfield  {journal} {\bibinfo  {journal} {Phys.
  Rev. E}\ }\textbf {\bibinfo {volume} {50}},\ \bibinfo {pages} {857} (\bibinfo
  {year} {1994})}\BibitemShut {NoStop}%
\bibitem [{\citenamefont {Kuhn}\ \emph {et~al.}(2007)\citenamefont {Kuhn},
  \citenamefont {Sigwarth}, \citenamefont {Miniatura}, \citenamefont
  {Delande},\ and\ \citenamefont {M{\"u}ller}}]{KUHN-2007-1}%
  \BibitemOpen
  \bibfield  {author} {\bibinfo {author} {\bibfnamefont {R.~C.}\ \bibnamefont
  {Kuhn}}, \bibinfo {author} {\bibfnamefont {O.}~\bibnamefont {Sigwarth}},
  \bibinfo {author} {\bibfnamefont {C.}~\bibnamefont {Miniatura}}, \bibinfo
  {author} {\bibfnamefont {D.}~\bibnamefont {Delande}}, \ and\ \bibinfo
  {author} {\bibfnamefont {C.~A.}\ \bibnamefont {M{\"u}ller}},\ }\href
  {\doibase 10.1088/1367-2630/9/6/161} {\bibfield  {journal} {\bibinfo
  {journal} {New J. Phys.}\ }\textbf {\bibinfo {volume} {9}},\ \bibinfo {pages}
  {161} (\bibinfo {year} {2007})}\BibitemShut {NoStop}%
\end{thebibliography}

%

\end{document}